\newcommand{\xmmn}{{\it XMM-Newton}}

\newcommand{\rxte}{{\it RXTE~\/}}

\newcommand{\iras}{{IRAS~13224-3809~\/}}

\newcommand{\rej}{RE~J1034+396}

\newcommand{\prd}{Phys. Rev. D}
\newcommand{\aap}{A\&A\/}

\newcommand{\aj}{AJ\/}
\newcommand{\apj}{ApJ\/}
\newcommand{\apjl}{ApJL\/} 
\newcommand{\apjs}{ApJS\/}
\newcommand{\araa}{ARA\&A\/}
\newcommand{\mnras}{MNRAS\/} 
\newcommand{\pasj}{PASJ\/} 

\newcommand{\nat}{Nature\/}
\newcommand{\aapr}{A\&AR\/}

\newcommand{\nar}{NewA Rev.\/}
\newcommand{\physrep}{Phys.~Rep.\/}

\newcommand{\na}{New Science\/}
\newcommand{\ssr}{Space Science Rev.\/}

\newcommand{\pasp}{PASP\/}

\def\Msun{\hbox{$M_{\odot}$}}

\def\sun{\hbox{$\rm {\odot}~$}}

\def\ergsec{\hbox{{$\rm ~erg~s^{-1}$}}}

\def\H0{{\rm ~km~s^{-1}~Mpc^{-1}}}

\def\d25{D$_{25}$}

\def\Mbh{\hbox{$M_{\rm BH}$}}

\def\mdotedd{\hbox{$\dot m_{\rm Edd}$}}

\def\Rg{\hbox{$R_{\rm g}$}}

\def\lsim{\mathrel{\hbox{\rlap{\hbox{\lower4pt\hbox{$\sim$}}}{\raise2pt\hbox{$<$}}}}}
\def\gsim{\mathrel{\hbox{\rlap{\hbox{\lower4pt\hbox{$\sim$}}}{\raise2pt\hbox{$>$}}}}}
\def\la{\mathrel{\hbox{\rlap{\hbox{\lower4pt\hbox{$\sim$}}}{\raise2pt\hbox{$<$}}}}}
\def\ga{\mathrel{\hbox{\rlap{\hbox{\lower4pt\hbox{$\sim$}}}{\raise2pt\hbox{$>$}}}}}

\newcommand{\be}{\begin{eqnarray}}
\newcommand{\ee}{\end{eqnarray}}

\documentclass[graybox]{svmult}

\usepackage{float}
\usepackage{mathptmx}        
\usepackage{amsmath}
\usepackage{amssymb}
\usepackage{color}
\usepackage{helvet}          
\usepackage{courier}         
\usepackage{dirtree}
\usepackage[normalem]{ulem}                             

\usepackage{makeidx}        
\usepackage{graphicx}        
\usepackage{subfig}

\usepackage{multicol}        
\usepackage[bottom]{footmisc}

\usepackage{multirow} 
\usepackage{hyperref} 

\usepackage[misc]{ifsym}
\usepackage{soul}
\usepackage{subfiles}       
\graphicspath{{./}{}}    

\begin{document}

\setcounter{chapter}{1}


\title{{The Super-Massive Black Hole close environment in Active Galactic Nuclei}}

\author{William Alston\thanks{European Space Agency (ESA), European Space Astronomy Center (ESAC), Villanueva de la Ca\~{n}ada, 28691 Madrid, Spain, \email{walston@sciops.esa.int}}, Margherita Giustini\thanks{Centro de Astrobiologia (CAB), CSIC-INTA, Villanueva de la Ca\~nada, 28692 Madrid, Spain, \email{mgiustini@cab.inta-csic.es}}, Pierre-Olivier Petrucci\thanks{Univ. Grenoble Alpes, CNRS, IPAG, 38000 Grenoble, France, , \email{pierre-olivier.petrucci@univ-grenoble-alpes.fr}}}

\maketitle

\abstract{Active Galactic Nuclei are powered by accretion of matter onto a supermassive black hole (SMBH) of mass $\Mbh \sim 10^{5-9} \Msun$.  The accretion process is indeed the most efficient mechanism for energy release we currently know of, with up to $\sim 30-40$\,\% of the gravitational rest mass energy that can be converted into radiation.  The vast majority of this energy is released at high energy (UV-X-rays) within the central $100$ gravitational radii from the central SMBH. This energy release occurs through a variety of emission and absorption mechanisms, spanning the entire electromagnetic spectrum.  The UV emission being commonly explained by the presence of an optically thick accretion flow, while the X-rays generally require a hotter, optically thinner, plasma, the so-called X-ray corona.  If outflows are present, they can also extract a significant part of the gravitational power.  With an origin in the deep potential well of the SMBH, the study of the high-energy emission of AGN give a direct insight into the physical properties of the accretion, ejection and radiative mechanisms occurring in the SMBH close environment. While not exhaustive, we discuss in this chapter our present understanding of these mechanisms, the limitations we are currently facing and the expected advances in the future.\\}

\section{Introduction}
\begin{figure}[t!]
\centering
\includegraphics[width=1.02\textwidth,angle=0]{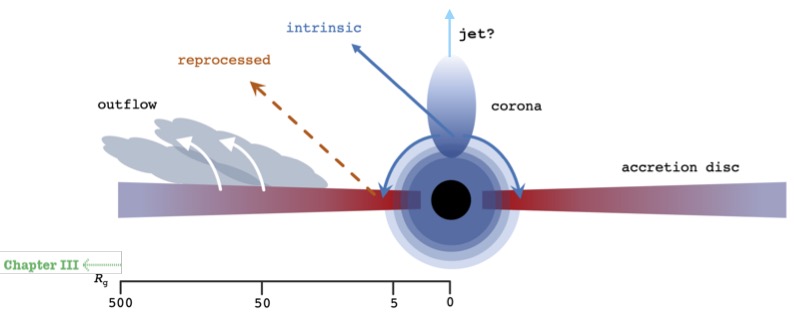}
\caption{\small Schematic cross section of the inner regions of AGN.  The accretion disc (red) extends down to a $\sim$~few~$\Rg$ near to the central BH.  The shape and size of the X-ray corona is still unknown, and is indicated as a spherical or lamppost geometry here (base of the jet?).  Its intrinsic power-law like emission (blue) irradiates the inner accretion flow and is reprocessed as the reflection spectrum.  In some sources, outflows are observed, which are believed to be launched from up to the inner $\sim 50 \Rg$. In this chapter we focus on $R < 100 R_g$, while winds produced on larger scales are discussed in Chapter 3.}
\label{fig:AGNsketch}
\end{figure}

There are strong evidences that the vast majority of galaxies harbor a supermassive black hole (SMBH) in their center, with a typical mass of $\Mbh \sim 10^{5-9} \Msun$. In about 10\% of these galaxies, a tremendous release of radiative power is observed. It is confined in the very inner region (size of the order of our solar system) of the host galaxy, in the vicinity of the SMBH. This region is called an Active Galactic Nucleus (AGN). The AGN emission can be of several orders of magnitude larger than the luminosity produced by the entire host galaxy.  AGN are believed to be powered by accretion of matter onto the central SMBH. Indeed, the accretion process is by far the most efficient mechanism for energy release we currently know of, with up to $\sim 40$\,\% of the gravitational rest mass energy that can be converted into radiation.  While in most of galaxies the SMBH is in a so-called quiescent state, in the case of AGN the SMBH is fed by a huge amount of matter coming from its close environment.  About $>$90\% of this energy is released within the central $100\, \Rg$ from the central SMBH, where $R_g$ is the gravitational radius\footnote{  $\displaystyle\Rg=\frac{G\Mbh}{c^2} \left(\simeq 1.5\times 10^{13} \frac{\Mbh}{10^8\Msun} \mbox{cm}\right )$, with $G$ the gravitational constant, $c$ the speed of light, and $\Msun$ the solar mass.}. This energy release occurs through a variety of emission and absorption mechanisms, covering the entire electromagnetic spectrum. 
It is generally thought to be dissipated partly in optical/UV as thermal heating in an optically thick ``cold'' plasma, the accretion disc, and partly in X-rays in a hot and optically thin plasma, the so-called hot corona (see Fig. \ref{fig:AGNsketch}). The accretion flow feeds the SMBH, while outflows of various kinds (wind and jets) may stream matter and energy away, extracting a significant part of the gravitational power. \\

In this chapter we discuss our present understanding of the close environment of SMBH in AGN, focusing on a few particular aspects which are presently debated in the X-ray astrophysics community. This includes the constraints on the hot corona properties (geometry, nature) in Sect. \ref{corona}, the important diagnostics coming from the emission and absorption processes occurring within the central $100\,\Rg$ 
and the inherent complexities of this environment in Sect. \ref{reprocessSect}. We also discuss in Sect. \ref{softXsect} the so-called Soft X-ray excess whose study has been revivified in the recent years. Then we present the different aspects of the high-energy variability in X-rays and Optical/UV in Sect. \ref{varSect} and the information that can be extracted from the multiwavelength correlations observed in sample of AGN in Sect. \ref{correlradioUVX}. We then conclude on the future prospects expected in these different domains in the coming years in Sect. \ref{futur}. \\ 


While most of the physical processes presented in the following sections are quite general and apply to the large majority of AGN, we mainly concentrate on nearby (primarily Seyfert galaxies) and moderately luminous AGN (i.e. with luminosity in the range $\sim 10^{-3}-10^{-1}$ $L_{Edd}$\footnote{The Eddington luminosity corresponds to the luminosity of an astrophysical object (such as a star) for which the radiation force compensates exactly the gravitational force. The Eddington luminosity depends only on the mass $M$ of the object: $L_{Edd}=4\pi GMm_p\frac{c}{\sigma_T}$, with $m_p$ the proton mass, $G$ the gravitational constant, $c$ the speed of light and $\sigma_T$ the Thomson cross section. $\displaystyle L_{Edd}\simeq 1.3\times 10^{46}\frac{M}{10^8\Msun}$ erg s$^{-1}$.}).

\section{The compact source of X-rays}
\label{corona}
X-ray emission makes up a large portion ($\sim 5-10$ per cent) of the AGN bolometric luminosity\footnote{The bolometric luminosity represents the total luminosity emitted by the AGN over all wavelengths.}, with $L_{\rm X} \sim 10^{40}-10^{44} \ergsec$.  The observed rapid variability  suggests the X-ray emission region is very compact, of the order of a fraction of a light hour\footnote{If the luminosity of an unresolved source varies significantly in a time scale $\Delta t$, then the radius of the source can be no larger than $R = c \Delta t$.}. The results from AGN gravitational microlensing measurements of quasars (e.g. \cite{morgan10,blackburne11,jimenez12}) 
have revealed insights into the size of the X-ray corona.  These at least tell us that the X-ray emitting regions are within the central $\sim 15$\, \Rg\ of the central BH.


Early observations found that the $2-20$ keV band photon flux $N(E)$ as a function of photon energy $E$ could be modelled with a power-law, $N(E)\propto E^{-\Gamma}$, of photon index $\Gamma \sim 1.9$ (e.g., \cite{mushotzky80,poundsetal90}).  This power-law continuum was later observed to have an exponential cut-off at $\sim 100$\,keV \cite{ZdziarskiETAL95} and, in the last ten years, this has been confirmed thanks to the high-energy capabilities of \textit{NuSTAR} which has allowed the properties of the hard X-ray emitting corona to be commonly explored up to $\sim~100$\,keV in several AGN \cite{fab15,tor18}.

The currently favoured model to explain this cut-off power-law shape for the primary continuum emission is the inverse--Comptonisation of optical/UV disc photons in an optically thin `corona' of hot (and possibly relativistic) electrons \cite{HaardtMaraschi91}. 
The electrons are assumed to be in thermal equilibrium and have a Maxwellian distribution with characteristic temperature $kT_{\rm e} \sim 100$\,keV ($T_{\rm e} \sim 10^9$\,K), which would explain the presence of the high energy cut-off discussed above, whilst the seed photons have initial energy $E \ll kT_{\rm seed} \sim 0.1$\,keV ($T_{\rm seed} \sim 10^5$\,K).
A spherical corona with a temperature $T_{\rm e}$, a typical Thomson depth, $\tau_{\rm T}\simeq 1$, and a radius $R_{10}$ (in units of $10\, R_{\rm g}$) above a BH of mass $m_6$ (in units of $10^6\,m_6\,\Msun$) has a very small mass of $M_{\rm cor}=10^{-8}\,m_6^2\,R_{10}^2\,\Msun$, a thermal energy $E_{\rm th}\approx 10^{42} \,m_6^2\,R_{10}^2$ erg and crossing time $t_{\rm cross}=50\,m_6\,R_{10}$ seconds. 
The luminosity obtained if all the coronal energy is released on a crossing time
$L_{\rm cross}\approx 2\times 10^{40}\,m_6\,R_{10}\ergsec$ is much less than the Eddington luminosity $L_{\rm Edd}$, in that $L_{\rm cross}/L_{\rm Edd}\approx 10^{-4}\,R_{10}$. Higher luminosities, as are typically observed, require that both cooling and heating occur on a faster timescale. 

While the release of the gravitational power is undoubtedly the source of the corona heating, how this heating is transferred to the particles and how particles reach thermal equilibrium is still not understood. The magnetic field is expected to play a major role (e.g. \cite{mer01}) but the details of the process are unknown. We even do not exactly know the true nature of the hot coronal plasma. It could be dominated by electron-positron pairs, the pair creation process playing the role of a natural thermostat to keep the plasma to a few hundreds of keV. This is an early idea put forward in order to provide such equilibrium (e.g. \cite{sve84}  
and see discussion in \cite{fab17}).  Here, the coronal plasma is assumed to be thermal and the temperature at the maximum allowed by electron-positron pair production: heating pushes the temperature upward until it is balanced by the creation of pairs. In the case where the hot corona is part of the accretion flow; however, a more standard electron-proton plasma is expected and the corona temperature results naturally from the radiative equilibrium between the hot corona and the accretion disc (e.g. \cite{HaardtMaraschi91}).


Different theoretical models have been proposed to explain the physical properties of this hot corona in the inner AGN regions (see e.g. \cite{yua14} for a review), between (as a non exhaustive list) Slim discs \cite{Abramowicz88}, Advection Dominated Accretion Flows (ADAF, \cite{NarayanYi95}), Adiabatic Inflow-Outflow Solutions (ADIOS, \cite{bla99}), Luminous Hot Accretion Flows (LHAF, \cite{yua01}, Magnetically Arrested Discs (MAD, \cite{narayan03}) or Jet Emitting Discs (JED, \cite{marcel18}). This hot flow may be formed in different ways, e.g., from disc evaporation \cite{mey00b} or Magneto-Hydro-Dynamic (hereafter MHD) accretion-ejection processes (e.g. \cite{fer06a,liska22}). But up to now, none of these models make a complete consensus and none of them is able to explain all the different high-energy properties of compact objects.

\begin{figure}
\centering
\includegraphics[width=.96\textwidth,angle=0]{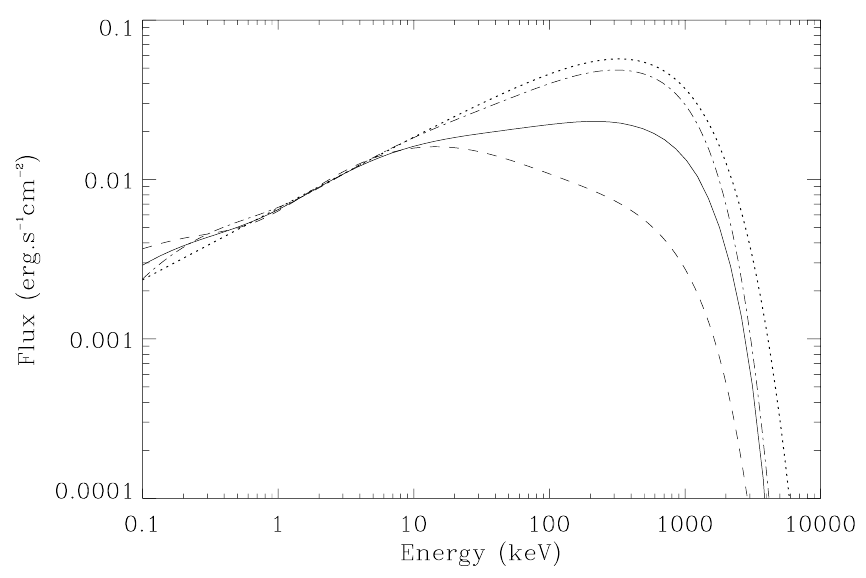}
\caption{\small Comptonised models for different hot corona geometries assuming an electron temperature $kT_e=360$ keV and a seed soft photon temperature $kT_{bb}=5$ eV. The slab, hemispherical, and spherical geometries are plotted in a solid, dashed, and dot-dashed line, respectively. The optical depths, chosen so as to
produce approximately the same spectral index for energies $<$ 10 keV, are 0.09,
0.16, and 0.33 for the slab, hemispherical, and spherical geometry, respectively. For comparison, we have also overplotted a cutoff power-law with a cut-off energy $E_c=2kT_e$
with a dotted line. From \cite{pet00}.}
\label{fig:degeneracy}
\end{figure}
It has been rapidly realized that it would be difficult anyway to constrain the main properties of the hot corona like e.g. its temperature, optical depth or the geometry with only X-ray spectroscopy if the energy range used to constrain the X-ray spectrum is too small. For example, different couples of temperature and optical depth values will produce similar Comptonisation spectra from different corona geometries (e.g. \cite{pet00} and see Fig. \ref{fig:degeneracy}). To break down this spectral degeneracy requires very broadband coverage in other parts of the high-energy spectrum, from the UV to the very hard (above 10 keV) X-rays. But X-ray polarimetry should also bring important constraints in this respect (see Sect. \ref{xraypolar}).


\section{Reprocessing of X-ray radiation in the gaseous environment close to the SMBH}\label{sec:reprocessing}
\label{reprocessSect}
The presence of substantial amounts of gas along the line of sight toward the luminous continuum emission source of AGN has been evident since early X-ray observations of both less luminous and local Seyfert galaxies (e.g., NGC 4151, \cite{1977MNRAS.181P..43B}) and more luminous and distant QSOs (e.g., MR 2251-178, \cite{1984ApJ...281...90H}). 
In fact, reprocessing features such as absorption edges and emission and absorption lines are commonly observed in the X-ray spectra of AGN, and allow to estimate the physical conditions of the gas responsible, together with the X-ray radiation impinging on it, for such features (e.g., \cite{1994MNRAS.268..405N,1997MNRAS.286..513R,2009A&ARv..17...47T}).
For example, variability of the absorption or emission spectral features (reprocessing features) on relatively short time scales implies a physical origin close to the source of emission for the gas responsible for its reprocessing.
In the regions very close to the central SMBH, general relativistic effects strongly shape the spacetime geometry itself, and observing reprocessing features created in such regions allows to measure fundamental physical quantities such as the spin of the SMBH \cite{2000PASP..112.1145F,2007ARA&A..45..441M,2014SSRv..183..277R}.

\subsection{Basics of X-ray photons interaction with matter}
X-ray photons can easily pierce through large column densities of gas $N_H=\int n(l)\, dl=\tau/\sigma_{T}$, where $n(l)$ is the particle number density along the line of sight, $\tau$ is the optical depth, and $\sigma_T$ is the Thomson cross-section.
Typical column densities observed in the X-ray spectra of AGN range from $10^{21}$ up to $> 10^{24}$ cm$^{-2}$. When the value $N_H=1/\sigma_T\sim 1.5\times 10^{24}$ cm$^{-2}$ is reached, then $\tau = 1$ and the Compton-thick regime is entered: the intrinsic flux of photons is severely attenuated by the intervening matter even for energies as large as a few keV \cite{1999NewA....4..191M}.

\begin{figure}[t!]
\centering
\includegraphics[width=.96\textwidth,angle=0]{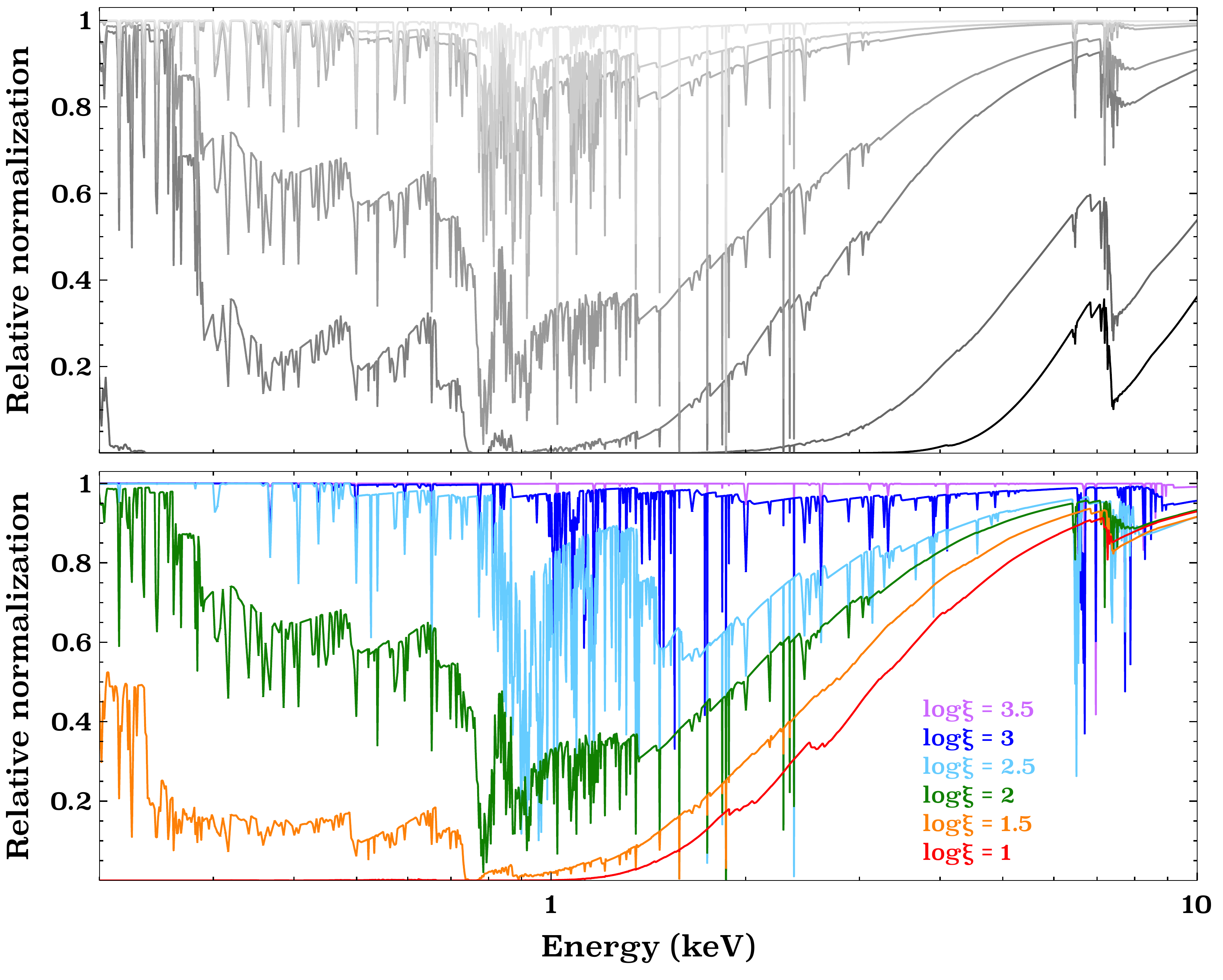}
\caption{\small Examples of X-ray absorption spectra for photoionized gas assuming a $\Gamma = 2$ power-law ionizing continuum. \textbf{Top panel:} $\log \xi = 2$
and different column densities $N_H$, from $10^{21}$ cm$^{-2}$ (uppermost, lightest line) to $10^{24}$ cm$^{-2}$ (lowest, darkest line). The photoelectric cutoff moves at higher energies with increasing column density; discrete absorption edges are visible, as well as many resonance absorption lines. \textbf{Bottom panel:} $N_H = 5\times 10^{22}$ cm$^{-2}$ and increasing ionization parameter $\xi$, from bottom to top. The gas becomes more transparent to the X-ray incoming flux with increasing ionization parameter: above $\log\xi\sim 3$, the broadband photoelectric cutoff is barely distinguishable from the continuum, and above $\log\xi\sim 3.5$ only absorption lines from the heaviest and most highly ionized atomic species are contributing to the absorption opacity.  All the models have been calculated with  \textsc{xstar} \cite{2001ApJS..133..221K}, which can be dowloaded from:  \url{https://heasarc.gsfc.nasa.gov/xstar/xstar.html}. }
\label{fig:absorption1}
\end{figure}

During their journey through such large gas column densities, the X-ray photons can be absorbed, leaving as a spectral signature the characteristic photoelectric cutoff that moves at higher energies for larger $N_H$, and absorption lines or edges when bound-bound or bound-free electron transitions are involved (Fig. \ref{fig:absorption1}, top panel). 

X-ray emission is also associated to the reprocessing, in particular fluorescence emission following K-shell absorption of the most abundant cosmic elements, notably iron (Fe K emission).
The Fe K fluorescence emission happens after an X-ray photon is absorbed by an iron atom of the reprocessing gas, ejecting an electron from the K-shell (which corresponds to the atomic energy level $n=1$), and the subsequent excited atomic state relaxes by fluorescence, i.e., an electron from the upper L-shell ($n=2$) moves to vacancy in the lower K-shell, releasing a photon\footnote{Alternatively, the excited atomic state can relax by emitting an Auger electron carrying 6.4 keV.} carrying 6.4 keV\footnote{There are actually multiple transitions associated to each atomic shell, with increasing energy and decreasing probability: the $n=2\rightarrow 1$ transition is called $K\alpha$, the $n=3\rightarrow 1$ transition is called $K\beta$, the $n=4\rightarrow 1$ transition is called $K\gamma$, ... while the $n=3\rightarrow 2$ transition is called $L\alpha$, the $n=4\rightarrow 2$ transition is called $L\beta$, and so on.}.
With increasing the ionization state of the iron atom, the energy emitted by fluorescence increases, to e.g., $E\sim 6.5-6.6$ keV for Fe XIX-XXIV, $E\sim 6.7$ keV for Fe XXV, $E\sim 6.97$ keV for Fe XXVI (e.g., \cite{2004ApJS..155..675K}).


The ionization state of photoionized gas is usually measured in terms of $\xi$, defined as the ratio of the ionizing photon flux density over the electron number density, which, for a spherically symmetric distribution of optically thin ($\tau < 1$) gas at a large distance $R$ from the ionizing photon source, can be expressed as $\xi=L_{ion}/4\pi n R^2$ where $L_{ion}$ is commonly the ionizing luminosity between 1 and 1000 Rydberg\footnote{One Rydberg=13.6 eV.}  \cite{1969ApJ...156..943T}.
The presence and the amount of matter between the AGN photon source of emission and the observer is inferred by the position and depth of the spectral curvature due to photoelectric absorption, as well of the absorption edges and the resonant absorption lines from ionized species, from C to Fe (Fig. \ref{fig:absorption1}, bottom panel).
Kinematic information about the gas responsible for the reprocessing features is also accessible through X-ray spectroscopy of good enough resolution to resolve the position of discrete emission or absorption lines. For example, if the absorption features are observed to be blueshifted with respect to the host galaxy cosmological redshift, the gas responsible for such features is inferred to be outflowing with a relevant velocity component along our line of sight; in the case of a relative redshift, the gas is inferred to be inflowing.
Synthetic X-ray absorption spectra contain simulations of the flux transmitted through optically thin column densities of gas, given an ionizing continuum; one-dimensional computations assuming geometrically thin gaseous slabs are implemented into photoionization codes as \textsc{xstar} \cite{2001ApJS..133..221K}, \textsc{cloudy} \cite{1998PASP..110..761F}. Examples of X-ray spectra of outflowing absorbing gas are shown in Chapter 3 of this Volume.

\subsection{X-ray reflection\label{sec:reflection}}

\begin{figure}[t!]
\centering
\includegraphics[width=.96\textwidth,angle=0]{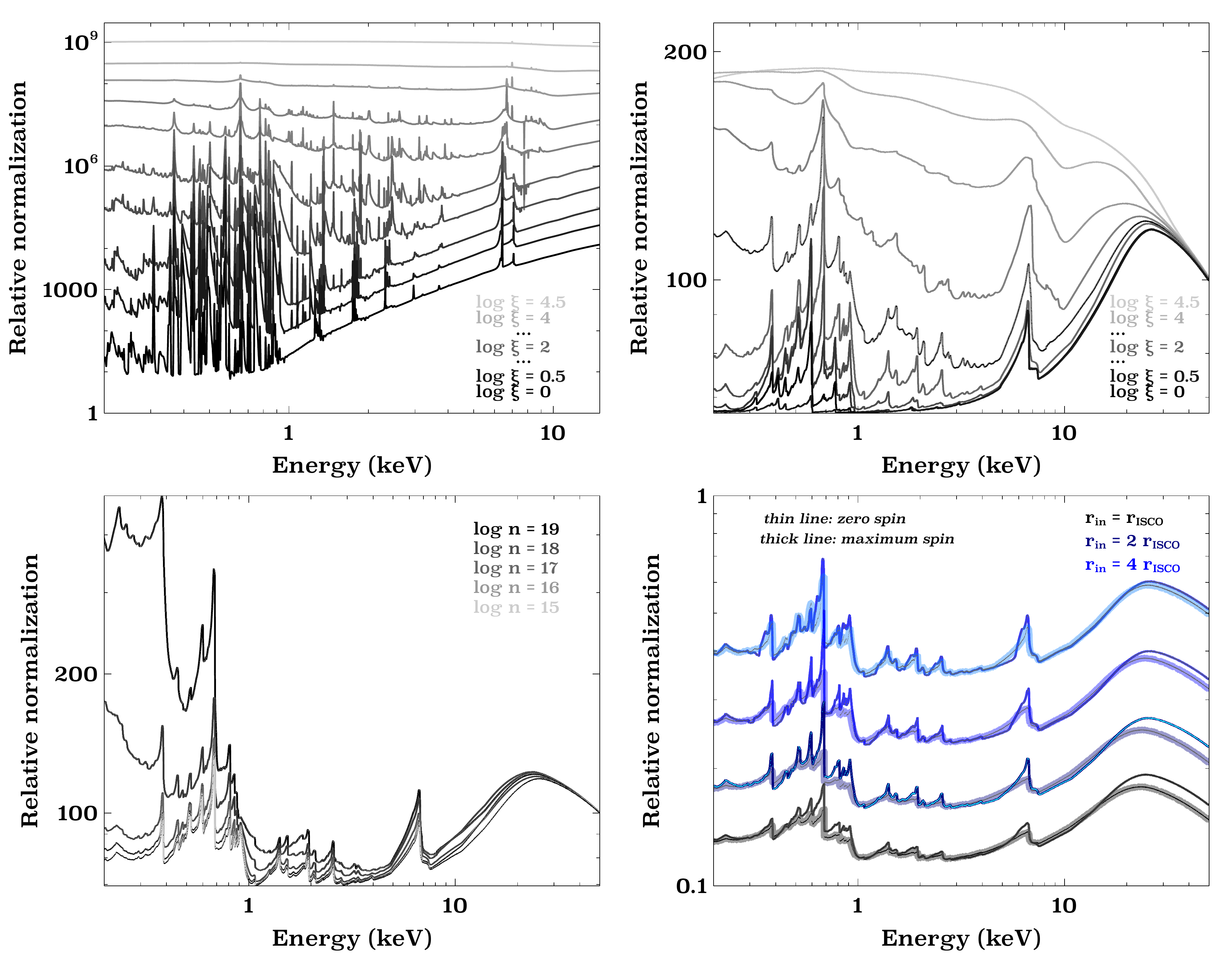}
\caption{\small Reflection spectra computed assuming a power-law with $\Gamma = 2$ illuminating a disc with solar abundances, emissivity index = -3 (i.e., the disc emissivity $\varepsilon(R)\propto R^{-3}$ at the radial distance $R$), inclined 30$^{\circ}$with respect to the line of sight. {\bf Top left:} non-relativistic reflection computed for a constant density  $n=10^{15}$ cm$^{-3}$ and various ionization parameters of the reflecting gas.
{\bf Top right:} relativistic reflection computed assuming a non-rotating BH and a disc with a fixed outer radius $r_{out} = 400 r_g$ and an inner radius $10\,r_{ISCO}$, with a constant density  $n=10^{15}$ cm$^{-3}$ and various ionisation states of the gas.
{\bf Bottom left:} relativistic reflection computed as above, but with a constant ionisation state  $\log \xi = 2$, and various densities of the reflecting gas.
{\bf Bottom right:} relativistic reflection  computed as above but for two values of the BH spin (zero or maximum) and various disc inner radii, from bottom to top and darkest to lightest: $r_{in} = 1$, 2, 4, and 8 times $r_{ISCO}$. For every case of $r_{in}$, the non-rotating BH case is plotted with a thin darker line, the maximally-rotating BH case with a thick lighter line. The four couples of models have been rescaled in normalization for visual purposes.\\ All the models have been calculated with \textsc{xillver} and \textsc{relxill}, which can be dowloaded from:  \url{http://www.sternwarte.uni-erlangen.de/~dauser/research/relxill/}. }
\label{fig:reflection1}
\end{figure}

The effects of reprocessing of the AGN X-ray continuum emission onto cold, optically thick matter (X-ray reflection) have been first considered by \cite{1988MNRAS.233..475G} and \cite{1988ApJ...335...57L}. The effects of photoelectric absorption, electron scattering, Comptonisation are to distort significantly the reprocessed X-ray spectral shape with respect to the simple illuminating power-law with $\Gamma \simeq 1.9$ generally observed. The most evident spectral features due to X-ray reflection are a broad excess of emission peaking in the $\sim 30-40$ keV range, the so-called Compton hump, and fluorescence emission lines of the most abundant cosmic metals, especially iron, which follow photoelectric absorption (e.g., \cite{1991A&A...247...25M,1991MNRAS.249..352G,1993MNRAS.261...74R}).

Modern simulations consider the effects of varying the ionization state or the density of the reflecting matter, and allow to test different geometries for the reprocessing matter configuration (e.g., \cite{1995MNRAS.273..837M,2005MNRAS.358..211R,2010ApJ...718..695G,2013ApJ...768..146G}; see Fig. \ref{fig:reflection1}). The possibility to also take into account different geometries of the X-ray continuum emission is crucial when modelling reprocessing features expected from matter located very close to the SMBH, where general relativistic effects are very strong \cite{2013MNRAS.430.1694D, 2014ApJ...782...76G}. Current state-of-the-art simulations merge the most extended reflection models, where the angular dependence of the radiation field is taken into account, with  general relativistic ray tracing of X-ray photons emitted from the surface of the accretion disc very close to the central SMBH, where strong field gravity effects are of crucial importance (e.g., \textsc{relxill},  \cite{2014ApJ...782...76G}, see the fourth panel of Fig. \ref{fig:reflection1}). 

In the strong field gravity regime (close to the SMBH, at $R < 10 \, R_g$), the reflected spectrum gets blurred by both special relativistic (Doppler broadening) and general relativistic effects (light bending and gravitational redshift).
In particular, the prominent Fe K emission line is expected to be broadened up to FWHM of $10,000$s of km~s$^{-1}$, and to become strongly asymmetric with a sharp ``blue wing'' and an extended ``red wing'' down to $E\sim 4-5$ keV \cite{1989MNRAS.238..729F,1991ApJ...376...90L,1993MNRAS.262..179M,2000PASP..112.1145F}. 
The strongest gravitational effects are expected close to maximally rotating BHs \cite{1996MNRAS.282L..53M}, where the radius of the innermost stable circular orbit (ISCO) shrinks to $R_{ISCO}\sim 1.2 \,R_g$ compared to the $R_{ISCO} = 6 \,R_g$ case for non-rotating BHs. 
When X-ray photons are emitted close to the ISCO around a maximally rotating SMBH, they produce a strongly blurred reflection spectrum where individual spectral features can be hard to discern, as they are almost completely smoothed out by general relativistic effects. In these cases, only the most prominent reprocessing features are expected to be visible against the intrinsic continuum emission, such as the Compton hump and Fe K and Fe L fluorescence emission.
It follows that by measuring accurately the spectral parameters of the broadened reflection features, of which the Fe K line is the easiest to measure for its prominence, the $R_{ISCO}$ can be inferred, and therefore the spin of the SMBH \cite{2006ApJ...652.1028B,2014MNRAS.444L.100D}.



\subsection{The fluorescent iron line}
\label{ironline}
The most common reprocessing feature observed in the X-ray spectra of AGN is the Fe K fluorescence emission line at $E\sim 6.4$ keV, with a moderate width and EW (FWHM$\sim$ 2000 km s$^{-1}$, EW$\sim 30-200$ eV; see e.g. \cite{2004ApJ...604...63Y,2011ApJ...738..147S,2014MNRAS.441.3622R}). Such feature is usually associated to distant reprocessing material, located at $R \gg 1000\, R_g$ (e.g., \cite{2006MNRAS.368L..62N}).
However, in many cases a much broader, stronger, and asymmetric Fe K emission line is observed, indicative, as explained in the previous section, of reprocessing happening very close to the central SMBH ($R \ll 100\, R_g$), where strong field gravity effects are important \cite{1989MNRAS.238..729F}.

Observations by {\it ASCA}  presented by \cite{1995Natur.375..659T} showed the first clear example of a very skewed Fe K emission line in the Seyfert galaxy MCG-6-30-15. The very broad line was inferred to be due to reprocessing in the very inner parts of the disc by an X-ray source of emission very compact and very close to the disc \cite{1995MNRAS.277L..11F}.

Early results based on {\it ASCA}  observations of a sample of Seyfert 1 galaxies showed evidences for relativistically blurred Fe K emission in about 3/4 of the sources \cite{1997ApJ...477..602N}.
Subsequent observations of an enlarged sample of Seyfert 1 galaxies with the more sensitive {\it XMM-Newton} satellite revealed the presence of relativistically broadened Fe K line in about a half of the sample; one third of the sample presented only narrow Fe K emission, while about one fourth of the sample displayed emission with an intermediate width, indicative of reprocessing at small, but not very small, distances from the SMBH \cite{2007MNRAS.382..194N}.

A systematic study of the {\it XMM-Newton} archive performed on about 150 type 1 AGN later confirmed the presence of relativistically broadened Fe K emission in one third of the sample \cite{2010A&A...524A..50D}, while a survey of about 50 AGN observed with {\it Suzaku} found the presence of relativistically blurred reflection in a half of the sample \cite{2012MNRAS.426.2522P}.

The temporal behaviour of the broadened Fe K line in the Seyfert galaxy MCG-6-30-15 also gave puzzling results to interpret \cite{1996MNRAS.282.1038I}, in particular uncorrelated variability of the Fe K line and continuum flux emission.
The puzzle can be solved assuming a very low height of the X-ray continuum emission source above the reflecting material, which is very close to ISCO around a maximally rotating BH; in this case, light bending due to the strong SMBH gravitational field 
explains the observed temporal and spectral properties of the broadened Fe K line in MCG-6-30-15 \cite{2002MNRAS.335L...1F,2003MNRAS.340L..28F,2004MNRAS.349.1435M}.

\begin{figure}[t!]
\centering
\includegraphics[width=.98\textwidth,angle=0]{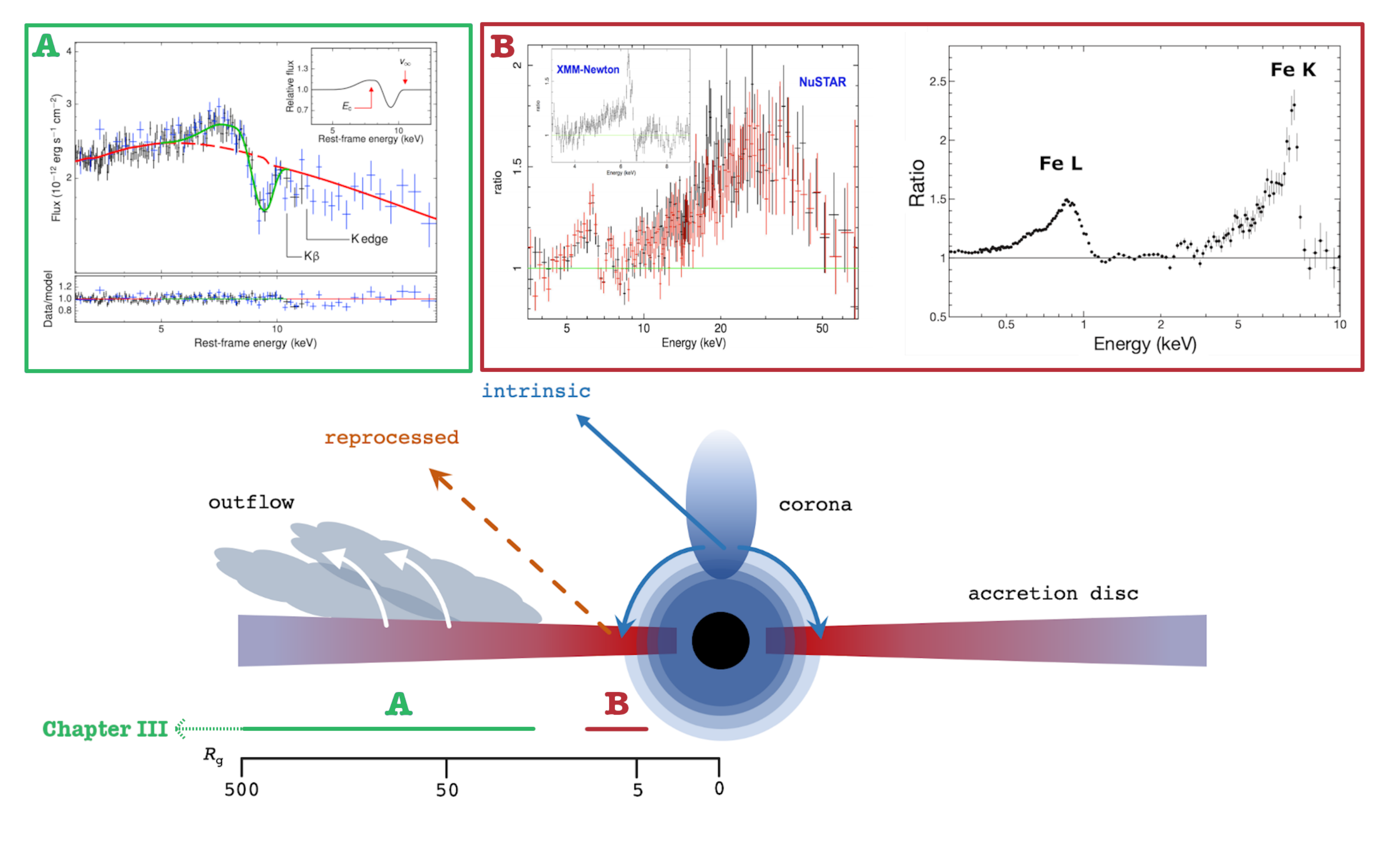}
\caption{\small  Panel A: best-fit model for the 4-20 keV spectrum of PDS 456 observed by XMM-Newton and NuSTAR, where the Fe K emission and absorption features have been modelled within an accretion disc wind scenario. The bottom panel shows data/model ratio, while the small inset shows the theoretical model applied; adapted from Figure 3 of \cite{2015Sci...347..860N}.  Panel B, left: NuSTAR data/model ratio of a power-law fit for the AGN NGC 1365, showing a strongly asymmetric Fe K emission line and a strong Compton hump; the small inset shows the simultaneous observation of the Fe K line with XMM-Newton. Adapted from Figure 1 of \cite{2013Natur.494..449R}. Panel B, right:  XMM-Newton data/model ratio of a power-law fit to the 1.5-3 keV band for the AGN 1H 0707-495, showing strong Fe L and Fe K emission lines which are strongly skewed and blurred by strong-gravity effects. Adapted from Figure 1 of \cite{2009Natur.459..540F}. Bottom panel: we have reported the schematic side view of the inner regions of AGN of Fig. \ref{fig:AGNsketch} with the origin of reprocessing features marked. The reprocessing features of panel A are produced between a few tens and few thousands $R_g$.  The reprocessing features of panel B are produced within the innermost few gravitational radii from the central SMBH, $R < 10\,R_g$.}
\label{fig:reflection2}
\end{figure}

In fact, independent of the time-variable nature of the Fe K line, in order to reproduce the very broad line profiles observed in Seyfert 1 galaxies, the so called lamp-post configuration needs usually to be invoked. Here the X-ray continuum source is assumed to be extremely compact, of the order of ISCO, and at a comparable distance from the SMBH (Region 'B' in the bottom panel of Figure~\ref{fig:reflection2}). The SMBH is maximally rotating, so that most of the energy is released very close to it and with a very steep emissivity index $k$, where the disc emissivity $\varepsilon(R)\propto R^{-k}$ \cite{2002A&A...383L..23M,2003MNRAS.344L..22M}.
The resulting line profile is very smooth and skewed toward the red.

The lamp-post model has been applied successfully to the high signal-to-noise ratio (S/N) broadband X-ray spectroscopic observations of several AGN, allowing to infer a very high spin for the majority of the sources studied \cite{2013MNRAS.428.2901W,2014SSRv..183..277R,2011MNRAS.416.2725P}. 
Broadband {\it XMM-Newton}  and {\it NuSTAR} observations of the Seyfert galaxy NGC 1365 revealed relativistically blurred features both in the Fe K band and at hard X-rays with a strong Compton hump peaking around 30-40 keV (\cite{2013Natur.494..449R}, Fig. \ref{fig:reflection2}, left in panel B).
Notably, in the Seyfert galaxy 1H 0707-495 both Fe K and Fe L emission lines were detected, and a highly-spinning BH was inferred (\cite{2009Natur.459..540F}, Fig. \ref{fig:reflection2}, right in panel B).

\subsection{Complex X-ray partial covering absorption\label{sec:pcov}}
The physical interpretation of the broad Fe K emission line is not completely unambiguous, as reprocessing features produced close to the SMBH, but not that close ($R\sim 100$s-$1000$s $R_g$), can give equivalent spectral results of relativistic reflection modelling.
If the source of X-ray emission has a non-negligible geometrical extension (a few $10\,R_g$) and the reprocessing material is not too distant from the continuum source of emission, then partially covering absorption can occur. 
Partially covering X-ray absorption is commonly detected at large distance from the SMBH (e.g., eclipses by Compton-thick clouds or by Compton-thin obscurers, see Chapter 1 of this Volume), but it might be relevant on scales of tens to hundreds of gravitational radii as well.

The archetypal broad Fe K emission line of MCG-6-30-15 can be reproduced in complex absorption scenarios, with layers of gas only partially covering the X-ray continuum emission source \cite{2008A&A...483..437M}, albeit with a contrived geometry \cite{2009MNRAS.397L..21R}. Broadband and deep X-ray observations  of MCG-6-30-15  performed with {\it NuSTAR}  and {\it XMM-Newton}  still did not allow to rule out the complex absorption scenario, although the relativistically blurred reflection is preferred on a statistical ground \cite{2014ApJ...787...83M}.
While a maximally rotating BH is almost invariably inferred to be in action in MCG-6-30-15 from spectral fits performed within a lamp-post model, this scenario is ruled out by \cite{2011MNRAS.416.2725P} who consider the effects of absorption in the Fe K band before modelling the relativistic emission line. 

X-ray partial covering can also explain the spectral curvature in the Fe K band  in 1H 0707-495 \cite{2004MNRAS.353.1064G,2004PASJ...56L...9T,2014PASJ...66..122M}, 1H 0419-577 \cite{2014A&A...563A..95D}, Mrk 335 \cite{2008ApJ...681..982G}, NGC 4151 \cite{2010ApJ...714.1497W}; all these AGN have been successfully modelled also in the blurred reflection scenario, e.g., by \cite{2012MNRAS.422.1914D,2005MNRAS.361..795F,2013MNRAS.428.1191G,2015ApJ...806..149K}. 
The physical context is therefore quite complex, and both relativistic reflection and partially covering absorption might be at play in shaping the appearance of X-ray spectra of AGN;
both components were requested to fit the X-ray spectra of e.g. NGC 4151  by \cite{2017A&A...603A..50B}, Mrk 335 by \cite{2019ApJ...875..150L}, and NGC 1365 by \cite{2009ApJ...696..160R}.
A way to break the degeneracy between the two scenarios is either using timing information (e.g., X-ray reverberation, Sect. \ref{reverb}) or high-resolution spectroscopy with the next-generation X-ray microcalorimeters (Sect. \ref{microcal}). 

\subsection{Reprocessing in the wind}
\label{sec:wind}
Further complexities in the interpretation of X-ray spectra of AGN were realised to exist after the investigation of the physical effects of outflowing, highly ionised matter on the incoming X-ray photon flux. 

During the past few decades, winds originating from the inner regions close to the SMBH have been recognised as a fundamental physical ingredient of AGN (see Chapters 1 and 3, this Volume; and \cite{2021NatAs...5...13L} for a recent review).
In particular, the work by \cite{2008MNRAS.388..611S,2010MNRAS.404.1369S} presented the first Monte Carlo simulations of X-ray radiative transfer through a simple biconical accretion disc wind, demonstrating that scattering and recombination into a wind launched at $R\sim 10$s-$100$s $R_g$ can produce a substantial excess of emission redward of 6 keV, together with strong absorption lines in the Fe K band, therefore confusing the simple interpretation of relativistically blurred Fe K emission.

\begin{figure}[t!]
\centering
\includegraphics[width=.8\textwidth,angle=0]{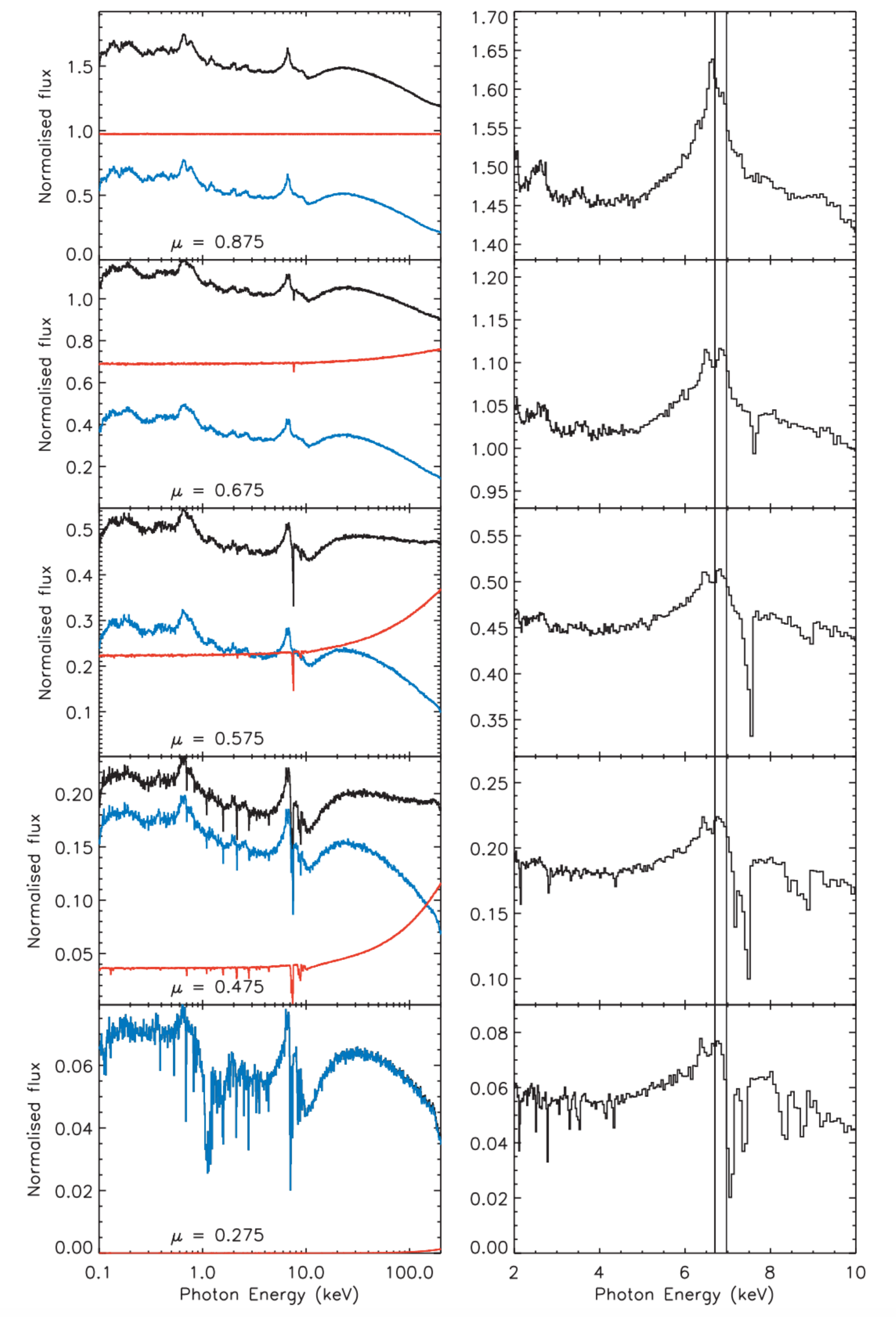}
\caption{\small Five examples of broadband 0.1-200 keV spectra expected in the disc wind scenario by \cite{2010MNRAS.404.1369S} for five different lines of sight, polar to equatorial from top to bottom ($\theta = \cos^{-1} \mu$ is the angle of the line of sight relative to the polar axis). The left panels report the 0.1-200 keV spectral components, the red line represents the transmitted spectrum, the blue line the scattered/reprocessed spectrum, and the black line the sum of the two. The right panels report a zoom in the Fe K band with the rest-frame energies of the Fe XXV/XXVI K$\alpha$ transitions ($\sim 6.7/6.97$ keV) marked. Note the different y-scale in the five panels. Figure 4 from \cite{2010MNRAS.404.1369S}.
}
\label{fig:absorption2}
\end{figure}

Figure \ref{fig:absorption2} shows five examples of broadband 0.1-200 keV spectra expected in the disc wind scenario by \cite{2010MNRAS.404.1369S} for five different lines of sight (l.o.s.). In the most equatorial view, the l.o.s. goes through the very dense base of the wind and the X-ray direct emission is completely suppressed. The spectrum is composed uniquely by the reprocessed component, which is dominated by strong absorption lines and broad emission features both in the soft and in the hard X-ray band, with red-skewed Fe L and Fe K emission lines and a strong bump of emission peaking around $30-40$ keV. 
In general, with decreasing the inclination angle between the polar axis and the l.o.s., the fraction of both transmitted and reprocessed flux increase, and less and less wind is intercepted by the l.o.s.; the absorption lines weakens and the relative importance of the scattered component increases. 
These trends were confirmed by \cite{2010MNRAS.408.1396S} who applied their Monte Carlo radiative transfer method to the output of hydrodynamic simulations of accretion disc winds presented by \cite{2004ApJ...616..688P}.

Observations of the luminous AGN in PDS 456, which shows a powerful, multiphase X-ray wind with strong reprocessing features \cite{2003ApJ...593L..65R}, revealed highly-variable X-ray absorbers on multiple time scales (\cite{2018ApJ...854L...8R}, see panel A of Fig. \ref{fig:reflection2}). 
A partially covering absorber 
was found to be variable on short time scales, placing it on radial scales of a couple of hundreds $R_g$ \cite{2016MNRAS.458.1311M}. That is close to the location where accretion disc winds can be launched, i.e., from a few $10 R_g$ for the fastest winds ($\upsilon_{out} > 0.3c$), to a few $100 R_g$ for the slower ones ($\upsilon_{out} < 0.1c$). 
Generally speaking, the closest the wind launching point to the SMBH is, the fastest the terminal velocity of the wind. Given the observed wind velocities in AGN, the inner launching radii for accretion disc winds can be therefore very small. For example, while the average components of the wind observed in PDS 456 have a velocity $\upsilon_{out}\sim 0.25c$ and are likely launched around $30-50\,R_g$, some of the fastest components with $\upsilon_{out}\sim 0.45c$ might arise from wind components launched as close as $\sim 10\,R_g$ from the central SMBH \cite{2018ApJ...854L...8R}.

\subsection{Strong-field gravity signatures in X-rays}
X-ray observations are a powerful tool to constrain the signatures of space-time distortions due to strong-field gravity effects.
One natural method to measure the SMBH spin is through X-ray reflection spectroscopy (Section~\ref{sec:reflection}), however, the physical complexities introduced in Section~\ref{sec:pcov} and \ref{sec:wind} have implications on the measurement of the spin of the SMBH.
While the application of the lamp-post model to the X-ray spectra of local Seyfert galaxies almost invariably reports a maximally spinning SMBH (e.g., Fig. 6 of \cite{2014SSRv..183..277R}),
the analysis of Suzaku observations of ``bare'' Seyfert, those AGN where the line of sight is free from substantial absorption ($N_H < 10^{22}$ cm$^{-2}$), revealed the presence of both partial covering and relativistic reflection, where maximally rotating BHs were not favoured by the data but rotating BHs with intermediate spins of about 0.7 were preferred \cite{2011MNRAS.411.2353P}. A variety of SMBH spin values was also found by \cite{2013MNRAS.428.2901W}, who analysed a sample of Seyfert galaxies selected to have a strong soft X-ray excess (see Section~\ref{softXsect}) and not to be dominated by absorption.
A recent thorough review of the complexities associated to the measurement of SMBH spin can be found in \cite{2020arXiv201108948R}.\\

Strong-field gravity effects have also been suggested through the presence of hot spots above the accretion disc. These hot spots are localized X-ray flares  which illuminate the accretion disc. They have been detected in at least four Seyfert galaxies: NGC 3516, where emission redward of the Fe K emission line, between 5.7 and 6.5 keV, was observed to be modulated and interpreted as flaring emission from the inner disc at about $10\,R_g$ \cite{2004MNRAS.355.1073I}; Mrk 766, where sinusoidal variations of Fe K emission are consistent with material orbiting at $\sim 100\,R_g$ from the central SMBH \cite{2006A&A...445...59T}; Ark 120, where flares redward and blueward of the Fe K emission line between 6 and 7 keV have been detected to vary on time scales of a few kiloseconds, placing the reprocessing material at about $10\,R_g$ \cite{2016ApJ...832...45N}; and NGC 2992, which hosts transient emission in the $5-7$ keV range which can be modelled with flares in the disc at $\sim 15-40\,R_g$ from the SMBH \cite{2020MNRAS.496.3412M}.
A systematic study of transient emission and absorption in the Fe K band of a large sample luminous Seyfert galaxies has been presented by \cite{2009A&A...507..159D}, who found evidence for relativistic, variable features to be around 30\%. 

The detailed study of the orbital motion of these hot spots could allow to measure the SMBH spin with future X-ray spectroscopic observations with large collecting area, which could allow a finely sliced time-resolved spectroscopy with high photon count statistic to be performed \cite{2004MNRAS.350..745D}. 




\section{The soft X-ray excess}
\label{softXsect}
\subsection{Observational signatures}
\begin{figure*}
\begin{center}
\includegraphics[width=0.9\textwidth]{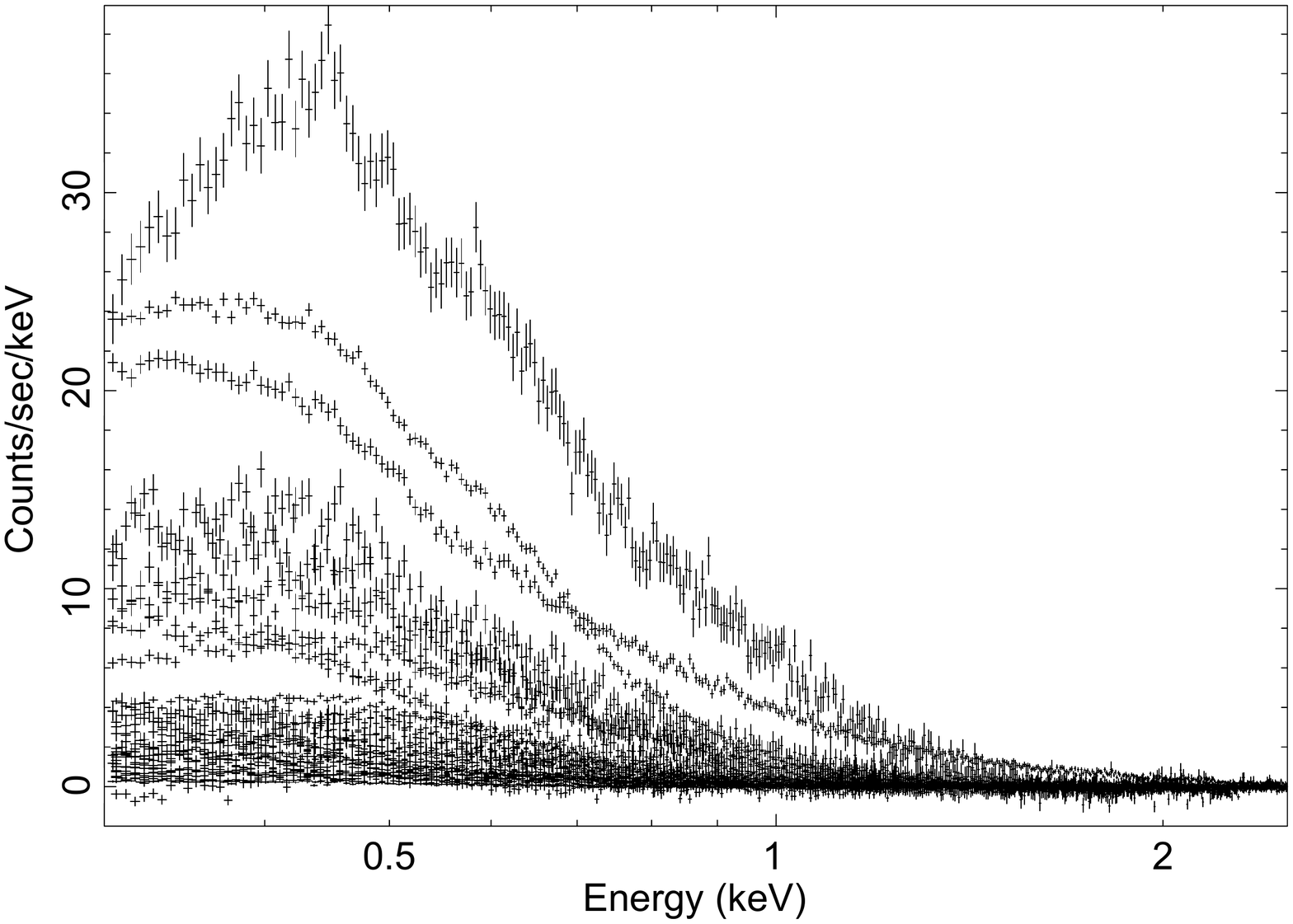}
\hspace*{-0.3cm}\includegraphics[width=0.97\textwidth]{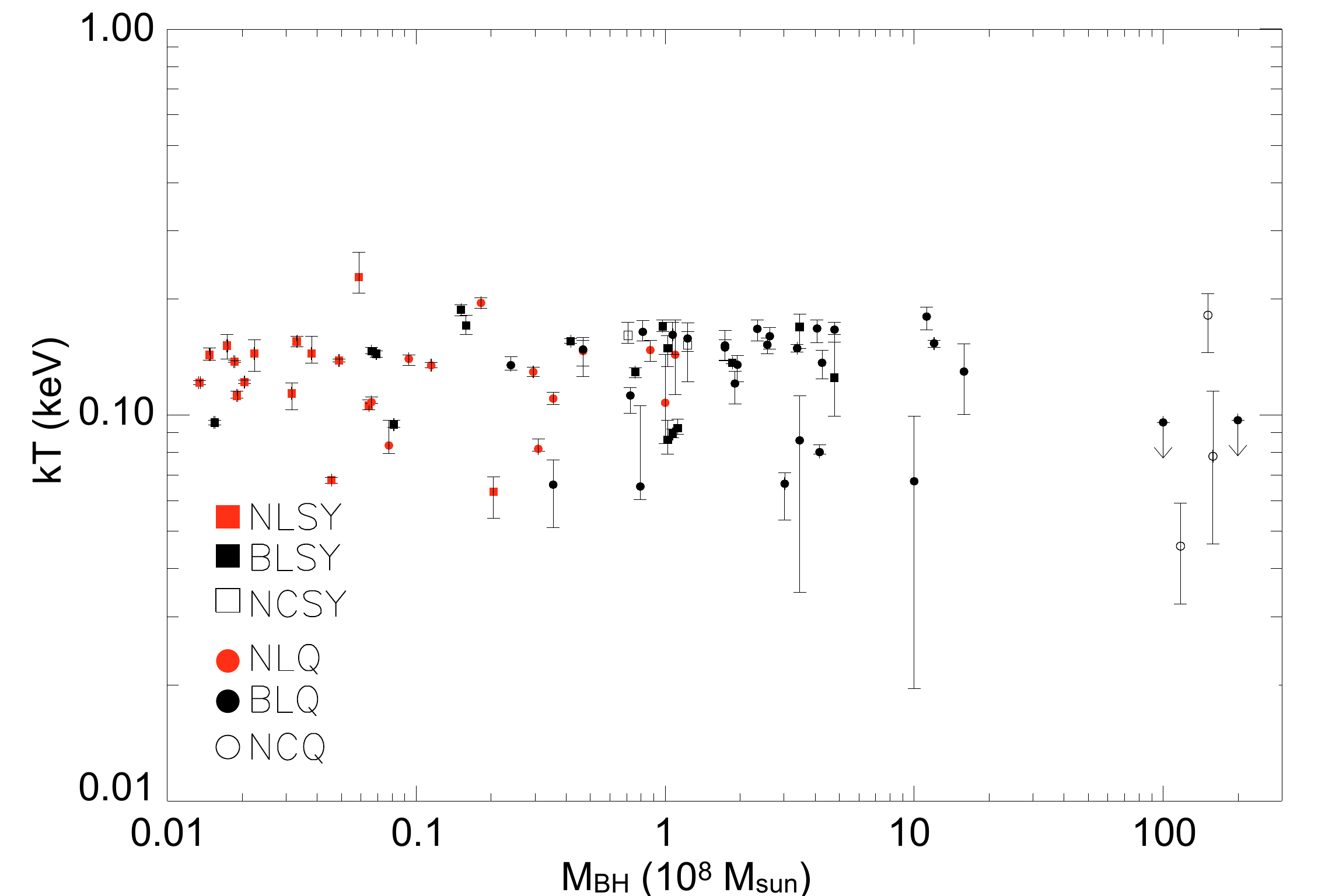}
\end{center}
\caption{{\bf Top:} Soft X-ray excesses in a sample of AGN (from \cite{cru06}). The sources were fit with a power-law for the primary continuum (i.e. to mimic Inverse Compton emission) and a blackbody for the soft X-ray excess. These are the residuals when the blackbody component is removed.  {\bf Bottom:} Temperature of the black body model, when used to model the soft excess, with respect to the BH mass (from \cite{bia09}).}
\label{fig:mkn841softX}
\end{figure*}

A large fraction of AGN show the presence of an excess of flux, below 2 keV, in comparison to the extrapolation of the power-law fit in the 3-10 keV range. 
 This excess is known as the soft X-ray excess. It was first discovered in the middle of the 80's  thanks to X-ray satellites with energy range covering the soft X-ray band ($<$2 keV) like {\it HEAO} or {\it EXOSAT} (e.g. \cite{arn85,sin85,tur89,wal93}). It was observed in several AGN by all the subsequent X-ray missions (e.g. {\it ROSAT}, {\it ASCA}, {\it BeppoSAX}) and is now commonly detected by {\it XMM-Newton}  or {\it Swift}  (see examples shown in Fig. \ref{fig:mkn841softX}). Actually it seems to be an ubiquitous component in AGN  (see e.g. \cite{gli20} and references therein) given that, to our knowledge, no study about AGN without soft X-ray excess have been published so far. Nevertheless, the physical origin of this component is yet still not understood.\\

The spectral shape of the soft X-ray excess does not show signatures of obvious atomic features (neither in emission nor absorption) and is generally well fit with a blackbody shape. It has been realized that the temperature $T_{bb}$ of the blackbody was always in the range 0.1-0.2 keV whatever the mass of the central BH or the luminosity of the AGN (e.g. \cite{gie04,por04,bia09}, see Fig. \ref{fig:mkn841softX} bottom). While the soft X-ray excess is widely believed to be related to the accretion disc, the energy range in which it is observed is incompatible with the expected temperature (of the order of a few eV) of a standard accretion disc around a  SMBH\footnote{The typical temperature of an accretion disc around a SMBH of mass $M_{BH}=M_8 10^8 M_{\sun}$ is approximately $T_{bb}\simeq 10^{5} M_8^{-1/4}$ K, e.g. \cite{ree84}.}. Moreover, the disc luminosity is expected to be $\propto T_{bb}^4$ 
which is also not observed (e.g. \cite{gie04,bia09}). These two arguments rule out the soft X-ray excess as the direct signature of the accretion disc emission. The remarkable constancy of the soft X-ray excess spectral shape across a large number of sources with different BH masses or accretion rates could suggest that it is not a ``true'' continuum component and that it is more related to micro-physics processes (e.g. atomic transitions) occurring at constant energies. Actually, both approaches, relying on a true continuum component or atomic processes, appear yet valuable as explained in the next section.

\begin{figure*}
\centering
\hspace*{-0.2cm}\includegraphics[width=0.9\textwidth,angle=0]{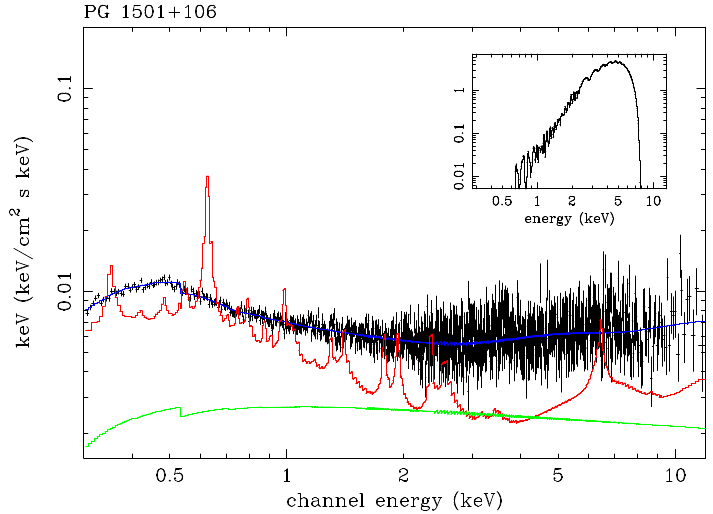}
\includegraphics[width=0.9\textwidth,angle=0]{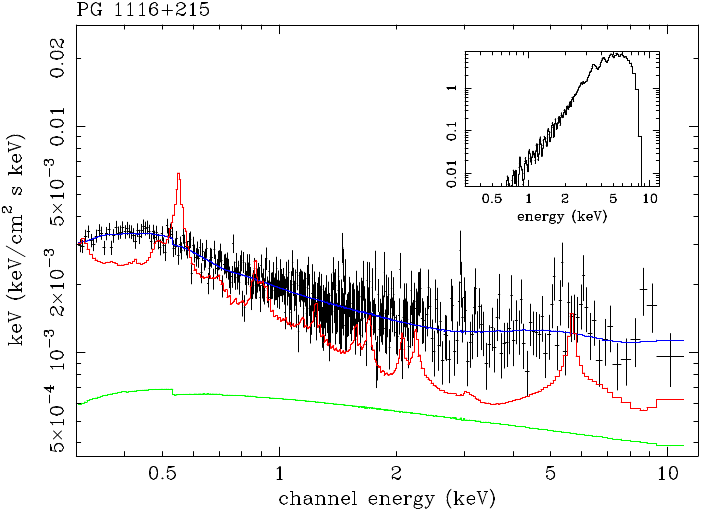}
\caption{Examples of best fit spectral energy distribution using relativistically-blurred ionized disc reflection models to reproduce the soft X-ray excess of two AGN. The black crosses are the data points, the blue line is the fit to the data, the green line is the power-law component of the fit, the red line is the reflection component of the fit, and the insert shows the relativistic line profile (from \cite{laor91}) used to blur the two components to produce the blue line model. The y axis has arbitrary units. From \cite{crummy06}.
}
\label{fig:blurredref}
\end{figure*}

\subsection{The soft X-ray excess modelling}
Several models have been proposed to explain the soft X-ray excess. The more discussed ones in the last 15-20 years are smeared absorption from partially ionized material (e.g. \cite{sch06}), blurred ionized reflection (e.g. 
\cite{cru06,jia19b}), and warm Comptonisation (e.g. 
\cite{mag98,don12,pet18}). 

The two first models rely on the presence of strong jumps in opacity between $\sim$0.5 and 2 keV caused by e.g. OVII/VIII and by the presence of several absorption/emission features from partially ionized ions like iron. The absence of emission/absorption features in the observed soft X-ray excess spectral shape requires then, in both cases, the presence of large velocity shears to smear out the (otherwise expected) sharp atomic features. This velocity shears could be produced by fast winds (for the "absorption" models) or strong gravity effects close to the BH (for "reflection" models) or a combination of the two. Precise modelling of disc winds absorption, however, have been shown to be unsuccessful in reproducing the smoothness of the soft X-ray spectral shape, the large number of atomic features expected across the 0.3-13 keV range being not sufficiently blurred by the too weak velocity shear of the wind (e.g. \cite{sch07}). 

%

\begin{figure}
\centering
\includegraphics[width=0.8\textwidth,angle=0]{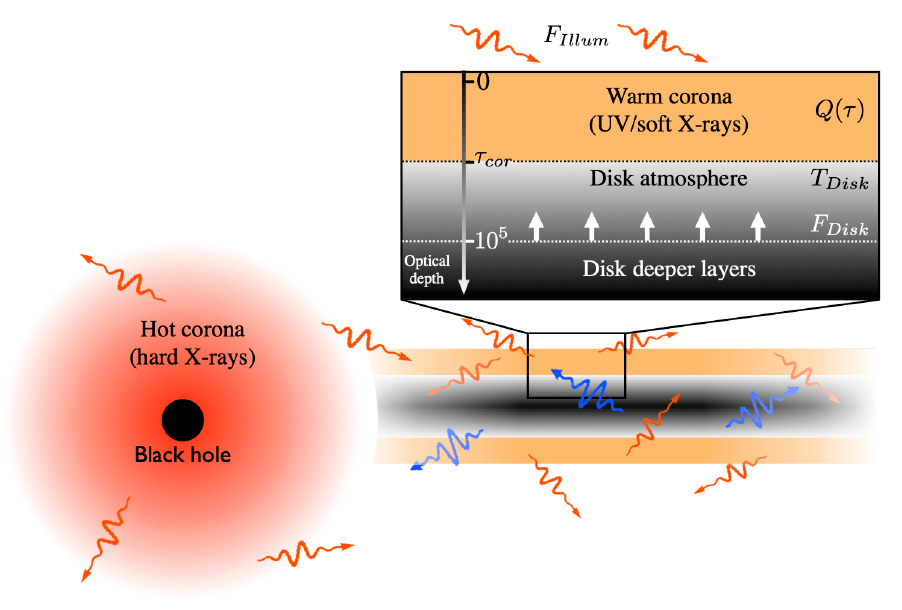}
\includegraphics[width=0.48\textwidth,height=5cm,angle=0]{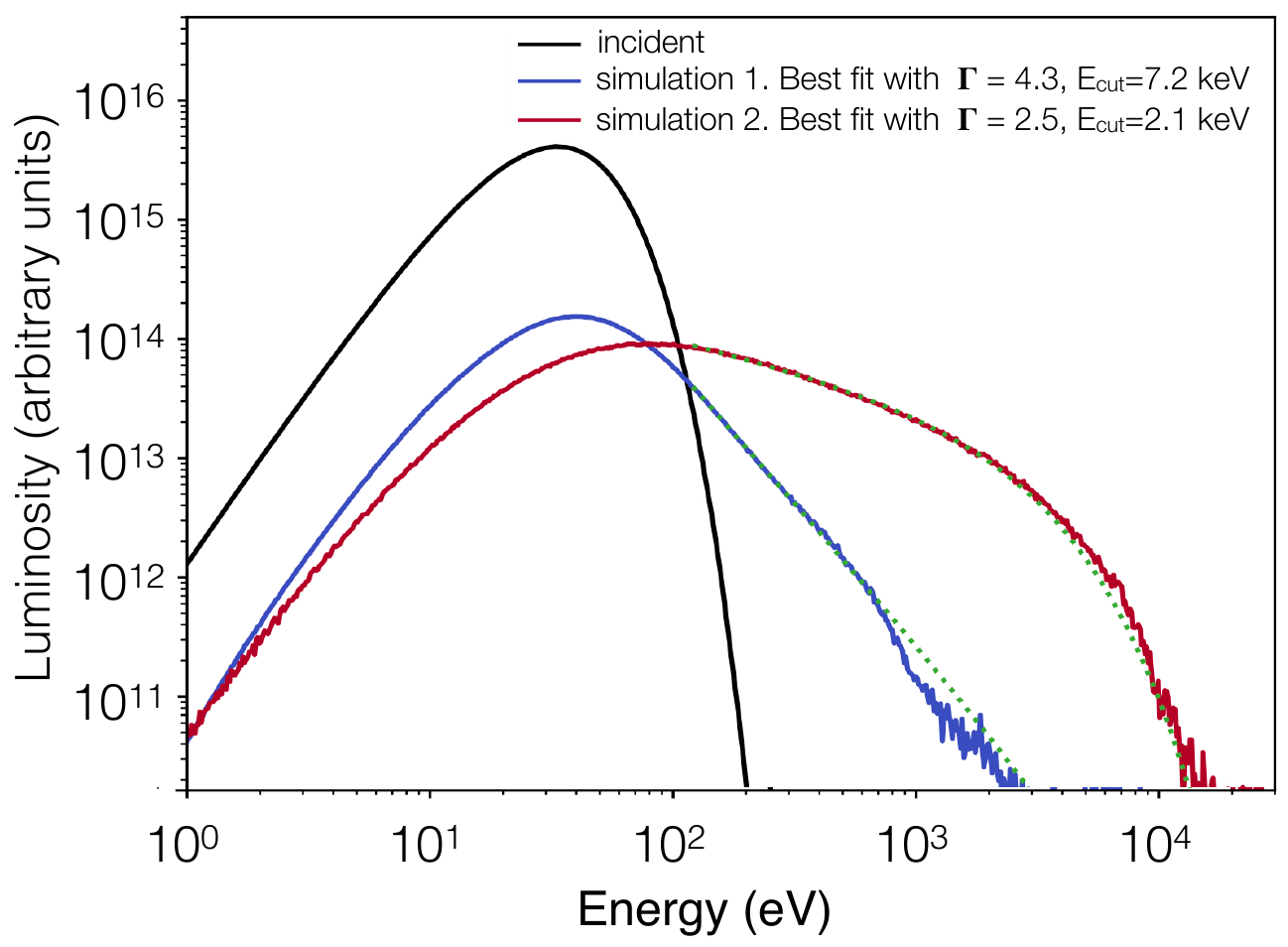}
\includegraphics[width=0.48\textwidth,height=5cm,angle=0]{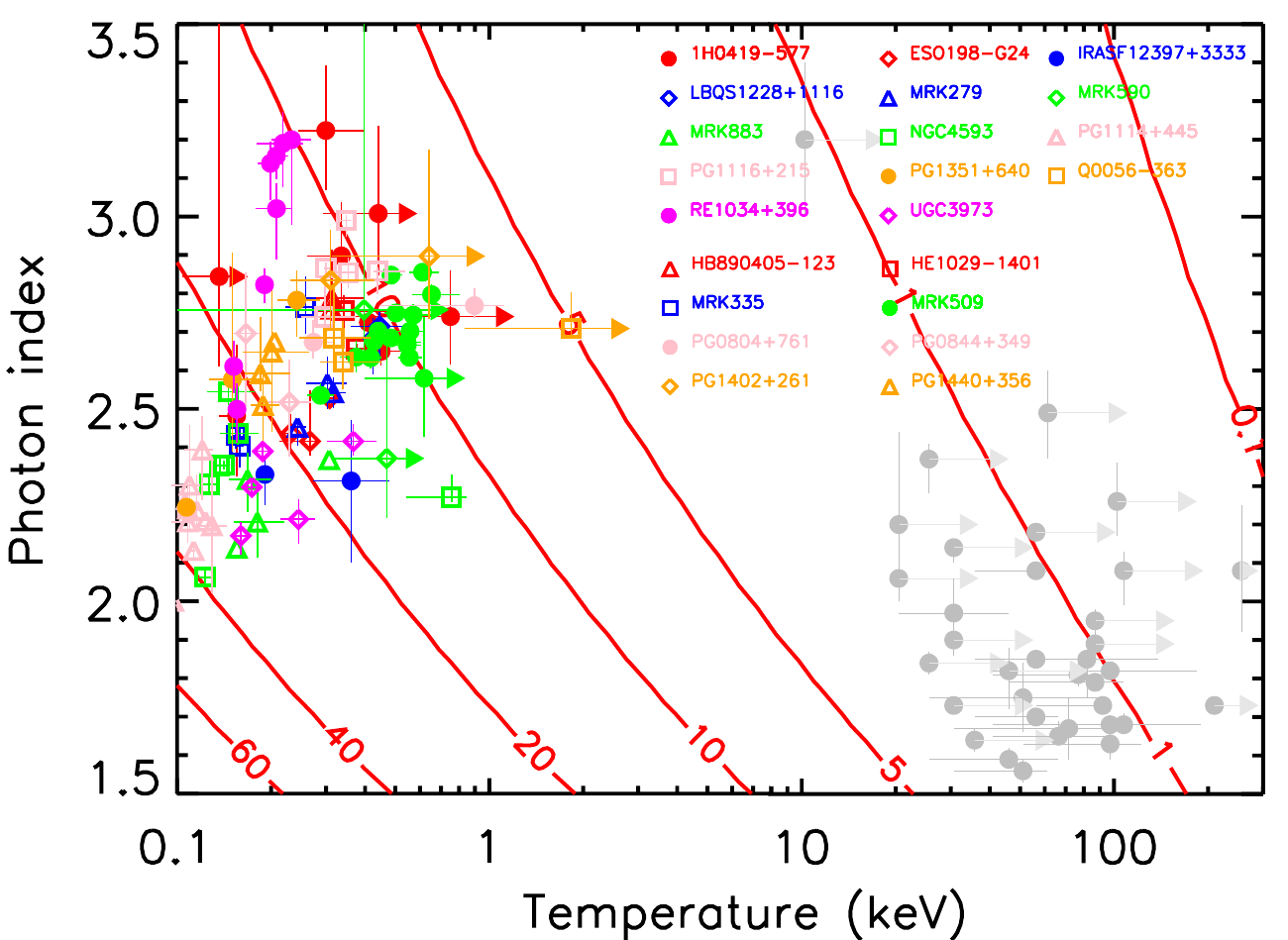}
\caption{{\bf Top:} Sketch of the two-coronae model. The hot ($kT_{hot} \sim 100$ keV) and optically thin ($\tau_{hot}\sim$1) corona  present in the inner parts of the accretion flow, close to the BH, is in red. The outer accretion flow is a vertically structured accretion disc (a zoom is shown in the figure) with cold and optically thick matter in the deeper layers (in black), characterized by a temperature $T_{Disc}$ and threaded by a vertical radiative flux $F_{Disc}$, while the upper layers (in orange) are composed of a warm ($kT_{warm}\sim$1 keV) and optically thick ($\tau_{warm}\sim$ 10-20) corona possessing a source of internal heating power $Q(\tau)$. The warm corona is also illuminated from above by an illumination flux $F_{Illum}$ produced by the hot corona. From \cite{pet20}. {\bf Bottom left:} Two examples of simulated spectra emitted by the warm corona for a corona optical depth of 20 and two different values of the internal heating (see \cite{pet20} for more details). {\bf Bottom right:} Best fit temperature and photon index of the warm corona in a sample of AGN, from \cite{pet18}. Different colored symbols are used for each object. They cluster in the left part of the figure, and agree with large ($>$5) optical depths (curved red lines). For comparison, the best fit photon indices of the hot corona (gray circles) from \cite{fab15} cluster in the bottom right part of the figure, where the optical depth is close to unity.}
\label{fig:warmcoronasketche}
\end{figure}
In blurred ionized reflection, the excess of emission in the soft band is due to a blend of emission lines present in the reflection spectrum. However, the reflection needs to be produced in the inner part of the accretion flow, close enough to the BH, where the relativistic effects are sufficiently strong to smear out these lines and reproduce the soft X-ray excess spectral shape (see Fig. \ref{fig:blurredref}). In addition to the soft X-ray excess, a blurred ionized reflection has the advantage of explaining also two other important spectral components commonly observed in AGN, i.e., broad iron lines and the Compton hump around 30 keV as dicussed in Sect. \ref{sec:reflection}. It has also been shown that the disc density can have significant effects on the excess of emission expected in the soft X-ray range by the reflection component (see e.g. Fig. \ref{fig:reflection1}). Indeed, the higher the density, the more efficient the free-free absorption processes which results in an increase of the temperature at the disc surface and then an increase of the reflected emission below 2 keV (e.g. \cite{gar16b,jia19b}).

An alternative scenario for the origin of the soft excess emission is the Warm Comptonisation model that assumes the presence of a warm ($\sim$ 1 keV) and optically thick ($\tau\gg$1) corona (different from the hot corona producing the hard X-rays), in the inner accretion flow, comptonizing the disc photons up to the soft X-ray range (e.g. \cite{CzernyElvis87,don12,pet13}). This model is also called the two-coronae model due to the presence of both warm and hot coronae to explain the UV/soft X-ray to hard X-ray emission (see e.g. \cite{meh11,urs20,mid20}).  The reproduction of the observed soft X-ray spectral shape requires temperatures and optical depth of the warm plasma to be in the ranges [0.1-1] keV and [10-30], respectively. More detailed radiative transfer analyses (e.g.  \cite{roz15}) show that the temperature of a few tenths of keV should be maintained in the entire warm corona, up to optical depth of a few tens, in order to have an emission dominated by Comptonisation and with a spectral shape in good agreement with observations.

Noticeably, the efficient Comptonisation inside the warm corona prevents the production of strong emission/absorption lines (e.g. \cite{pet20,bal20}) in agreement with observations. These results question, however, the origin of this warm corona. It has been proposed that the warm corona could be the inner part of the optically thick accretion flow, connecting the accretion disc to the hot inner corona (e.g. \cite{don12}), or the upper layers of the accretion disc (e.g. \cite{pet13}, see Fig. \ref{fig:warmcoronasketche}). The similar temperature and optical depth observed in sample of AGN (e.g. \cite{jin12a,pet18}) require, however, some types of thermostat that would provide the same warm corona properties whatever the AGN BH masses and disc accretion rates. Actually, this would imply some local (magnetic?) heating deposit across the entire warm corona up to large optical depth ($\gg 1$, see \cite{gro20,pet20,bal20}). This heating cannot be simply provided by the external illumination from the hot corona, since the corresponding heating deposit would occur only at the surface of the disc up to optical depths of a few.

\section{X-ray and Optical/UV variability}
\label{varSect}
\subsection{Aperiodic variability}
Active galaxies display aperiodic variability in all wavebands and on all timescales probed so far (e.g. \cite{Gaskell06}).  Strong UV and X-ray variability is common to AGN on a wide range of timescales with the most rapid variations seen in X-rays (e.g. \cite{mushotzky93}, \cite{alston13a,alston19a}).  The observed light curve, or time series, from an accreting source follows a stochastic (random) noise process.  Each recorded light curve from an individual source is one realization of the same underlying stochastic process (e.g. \cite{priestley81}; for an overview of analysis methods used on AGN light curves see e.g. Chapter 7 of \cite{bambi2020tutorial}).  Understanding this stochastic process provides additional and complementary insights to spectral and spatial information.\\

Over the last 30 years, the \rxte and \xmmn\ satellites have dramatically improved our understanding of the X-ray variability from accreting systems.  It too became clear that the variability process in AGN have much in common with X-ray Binaries (XRBs), scaled by mass and \mdotedd\ (see e.g. \cite{mchardy10rev}). 
This means that understanding the variability process in one class of objects can aid our understanding in the other.  BH-XRBs have a higher flux and higher count rates, making observations typically less affected by photon counting noise.  However, more photons per \textit{characteristic timescale} are typically received from nearby AGN.  This allows us to probe the variability at even faster characteristic timescales in AGN: timescales associated with the closest radii to the central SMBH.\\

\begin{figure}[t!]
\centering
\includegraphics[width=0.78\textwidth]{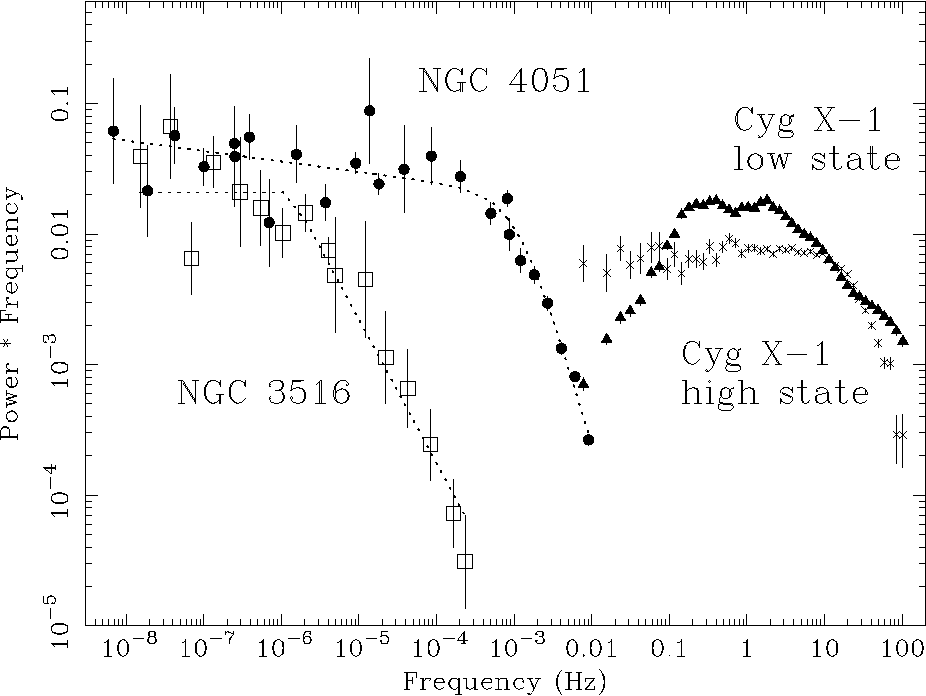}\\
\includegraphics[width=0.79\textwidth]{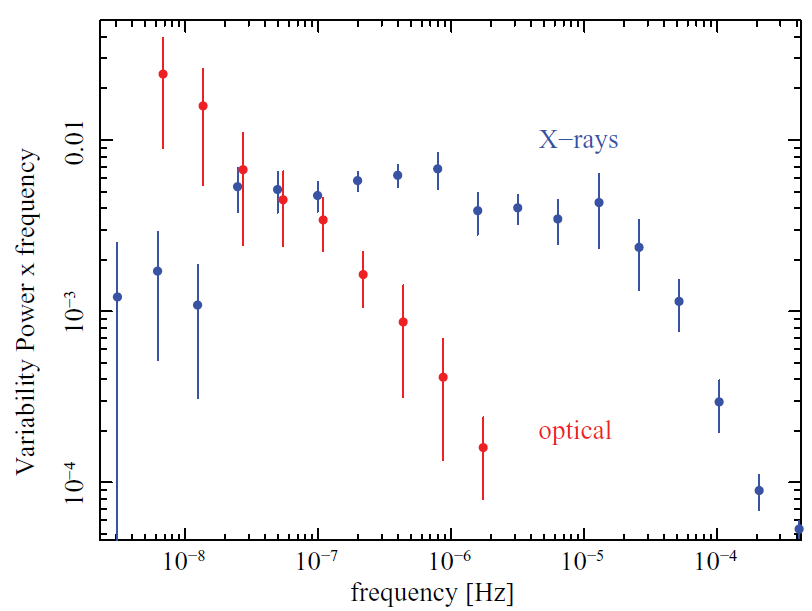}
\caption[X-ray timing properties of accreting BHs]{Top: Power spectral density of accreting BH systems, spanning 10 decades in frequency (from \cite{mchardy04}).  Cyg X-1 is a stellar mass BH-XRB with $\Mbh \sim 10 \Msun$ and NGC 4051 is a Seyfert 1 with a $\Mbh \sim 2 \times 10^6 \Msun$.  Bottom: Optical and X-ray PSD for the Seyfert 1 galaxy NGC 3783 \cite{arevalo13}.  At high frequencies, the X-ray variability power dominates, whereas at lower frequencies more power is seen in the optical.}
\label{fig:psd}
\end{figure}

XRBs are observed to transition through `states' on timescales of weeks to years.  States are characterised by their spectral hardness, $0.2-10$\,keV X-ray flux as well as observed variability properties (see \cite{belloni10rev} and \cite{fender10rev} for a review).  The most commonly observed states are the low/hard (LH), high/soft (HS) and the very high/intermediate (VHS), which represent different changing \mdotedd\ as the system evolves through an outburst.  The variability properties also change between the different states, and may offer a cleaner way of determining the state of the accreting BH (see e.g. \cite{mchardy10rev}).  If state transitions occur in AGN, the timescale will be much longer, of the order of $\sim$thousands of years for $\Mbh \sim 10^{6} \Msun$.  A statistical analysis is therefore required to understand AGN states.
\\


\noindent A variety of methods for characterising the variability are available to an observer in both the \emph{time} and \emph{Fourier} domain.  One Fourier method routinely applied to XRBs and AGN is the power spectral density (PSD) of the light curve (see e.g. \cite{vaughan03a}.  This provides information about the variability amplitude as a function of Fourier frequency, which is $1/timescale$.  An example of AGN and BH-XRB PSD are shown in Fig.~\ref{fig:psd}, where the similarities in PSD shape between the two classes of objects can be seen.  The PSDs in most AGN show a high frequency bend, which is similar to what is observed in  the soft state of XRBs.  The variability is that of broadband noise with the PSD shape well described by a bending power-law, $P(f) \propto f^{- \alpha}$ with $\alpha \sim 2$ at frequencies above the break $\nu_{\rm B}$, and $\alpha \sim 1$ below $\nu_{\rm B}$ (e.g. \cite{GonzalezVaughan13}).  The high frequency slope is commonly referred to as the ``red noise"  slope of the PSD, where the variability amplitude decreases with increasing Fourier frequency. The resulting light curve of such a PSD shape closely resembles a random walk.

The PSD is not always well described by a simple power-law or bending power-law model, with some objects displaying a `clumpy' or peaked structure (e.g. Ark 564; \cite{mchardy07} and \iras\; \cite{alston19a}).  In these two sources, the PSD was well described by multiple broad Lorentzian profiles.  These observations, together with the hardening of the PSD with energy at high frequencies, supports the idea of two variability processes, with one dominating at high and the other at low frequencies.\\

An important diagnostic of the underlying variability process is the linear rms-flux relation (e.g. \cite{UttleyMchardy01,uttleymchardyvaughan05}). Here, a linear relationship is observed between the root-mean-square (rms) amplitude and the mean X-ray flux over a given time light curve interval.  This linear relation appears to hold for all variability timescales measured; the timescale being probed can be changed by changing the length or bin width of the light curve, equivalent to integrating under different regions of the PSD (see e.g. \cite{vaughan03a} for more details).

This relation implies the model describing the underlying noise process is non-linear, for example an exponentiation of an underlying Gaussian distribution of variations, and predicts a log-normal distribution of fluxes.  The linear rms-flux relation and non-linear behaviour means the model must be multiplicative (rather than additive) and couple on all time scales \cite{uttleymchardyvaughan05,vaughanuttley07,alston19b}, and rules out certain types of model for the origin of the variability.



The leading model for the observed X-ray variability in AGN (like that in XRBs) is based on the inward propagation of random accretion rate fluctuations coupled at each radius in an accretion disc (e.g. \cite{lyubarskii97,kotov01,king04,zdziarski05,arevalouttley06,ingramdone10,kelly11}).  In this model, the local mass accretion rate through the inner regions of the disc modulates the X-ray emission.  This multiplicative model reproduces many of the currently known spectral variability patterns in AGN and XRBs \cite{arevalouttley06}, e.g. the linear rms-flux relation; the change in PSD shape with energy due to the radial dependence on emissivity; PSD break frequency dependence on energy; and observed energy dependent time lags.\\


\subsection{X-ray reverberation mapping}
\label{reverb}

\begin{figure}[t!]
\centering
\includegraphics[width=.99\textwidth,angle=0]{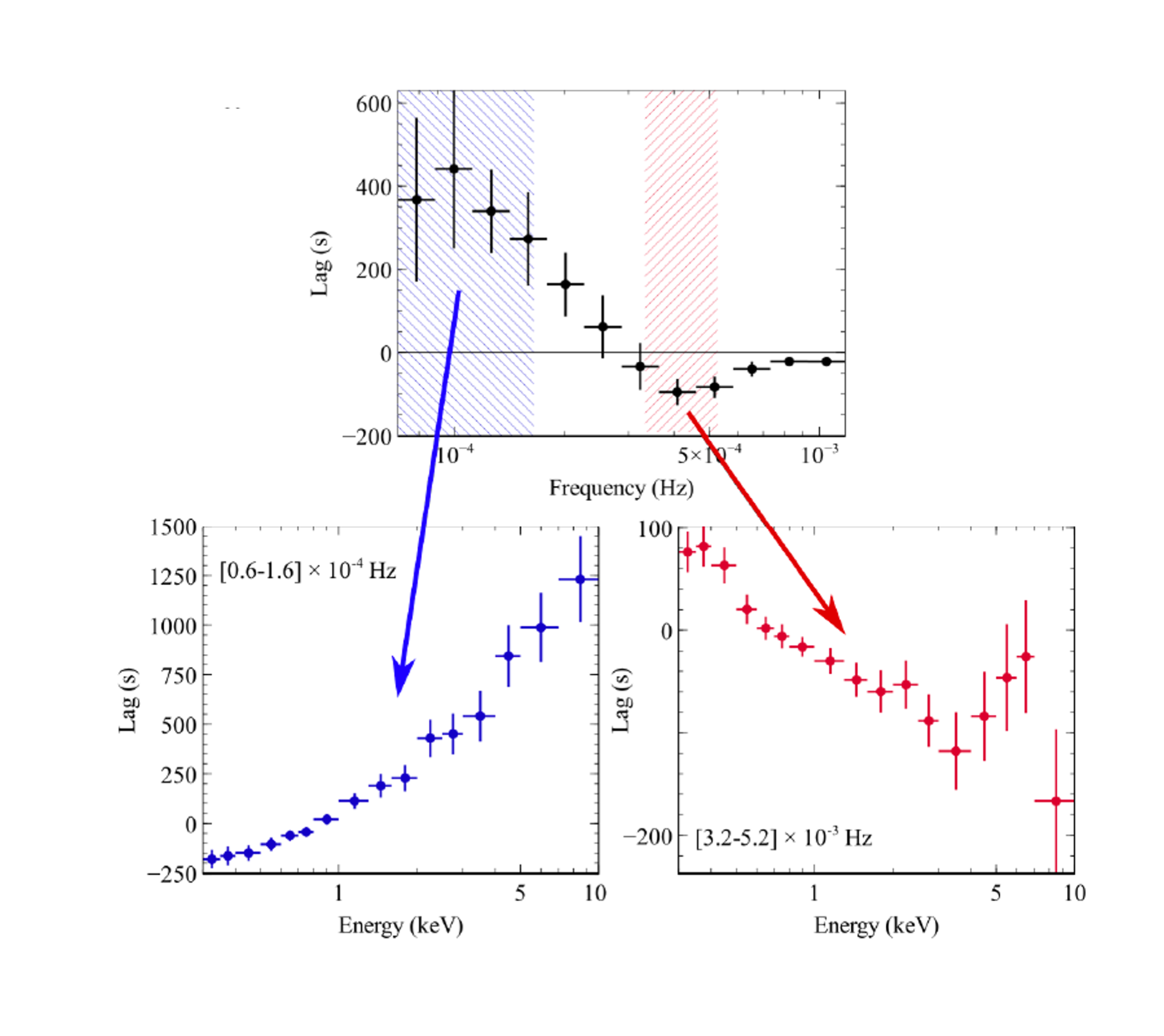}
\caption{\small Observed time delays in the active galaxy Ark 564, from \cite{uttley14rev}. The top panel shows the time lag as a function of Fourier frequency between a hard and soft X-ray band.  A positive lag value means the hard band variations follow that of the soft band, and vice versa.  The bottom panels show the energy-dependent time lags taken from two broad frequency ranges.  The left panel shows the low-frequency hard (intrinsic) lags (blue) and right panel shows the high-frequency soft lags (red), also refereed to as the reverberation signal.
}
\label{fig:reverbpica}
\end{figure}

\begin{figure}[t!]
\centering
\includegraphics[width=.99\textwidth,angle=0]{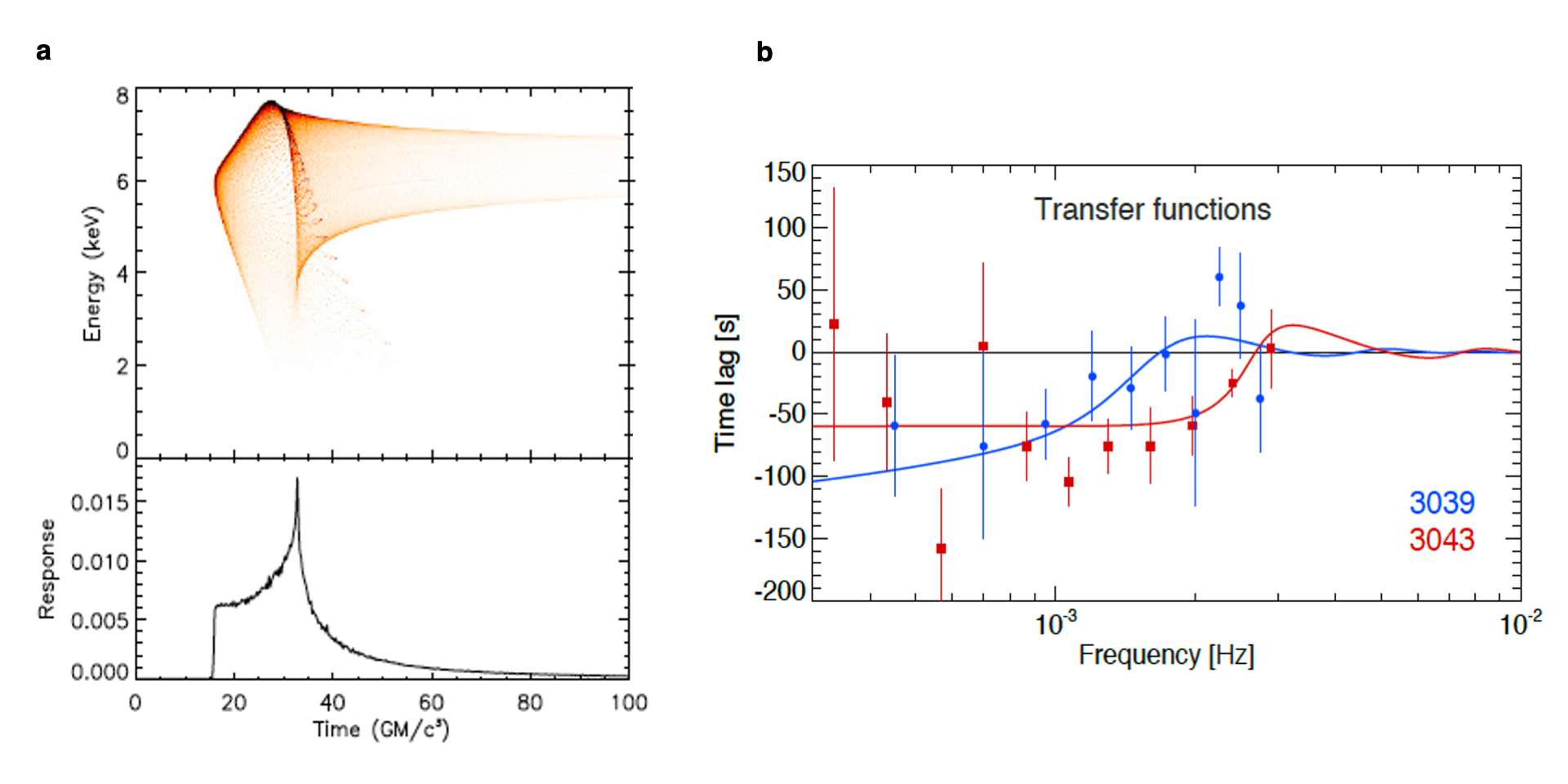}
\caption{\small \textbf{a.}  The energy dependent response of the Fe K$\alpha$ in the accretion disc to an impulse of radiation from the corona \cite{wilkins12}.  The path length between these components as well as the location of the inner disc sets the reverberation timescale.  GR effects mean the line response shifts to different energies as the flash of coronal emission evolves over the disc.
\textbf{b.} The Fourier transform of the disc response gives the ``transfer function'' which is modelled to the Fourier dependent time lags.  Shown are two models with different corona height above the disc in two observations of the NLS1, IRAS~13224--3809. The larger source height (blue) and hence larger path length has the first zero crossing at lower Fourier frequency  (\cite{alston20}).
}
\label{fig:reverbpicb}
\end{figure}

One of the most significant recent discoveries in accretion physics, is the detection of high-frequency X-ray \emph{reverberation} echoes in nearby NLS1s \cite{fabian09}.  These are observed as very short ($10-100$\,s) time delays between the primary X-ray emission (typically $1.0-4.0$\,keV) and the soft-excess-dominated band (typically $0.3 - 1.0$\,keV).  The reverberation signal is detected at high Fourier frequencies (short timescales), typically $10^{-4} - 10^{-3}$\,Hz.  This short timescale implies these signals originate at very short radii - ($\lsim 15$\,\Rg) - in the strongly distorted spacetime around accreting SMBHs \cite{demarco13lags,kara16}.  This is a region around the SMBH which cannot yet be directly spatially resolved in AGN by any telescope (the Event Horizon Telescope will only image the closest $\sim 2$ very-low accretion rate objects).   

A coherent picture is emerging where this reverberation signal is produced when the intrinsic coronal X-ray emission is reprocessed in the inner accretion disc: the \emph{timing} signature of the reflection spectrum, e.g. \cite{zoghbi12a, uttley14rev}) - see Fig.~\ref{fig:reverbpica}).   This signal encodes all the geometry, gravitational light bending and relativistic effects, providing a direct timing-based measure of mass and spin, as well as details of the accretion flow.  This information can be retrieved from the observed time lags using ``transfer functions''.  These model the response of the disc to a flash of X-rays coming from the corona - allowing us to decode the physical size scale of the accretion geometry and location of the emission processes (Fig.~\ref{fig:reverbpicb}).  The advantage is that reverberation provides information in absolute physical units, unlike the energy spectrum which provides estimates in the scale units of $\Rg$: with X-ray reverberation, when we measure the geometry we measure both the mass \textit{and} spin.  These results also suggest that reflection emission is part of the soft X-ray excess observed in AGN (see Sect.~\ref{softXsect}).  These signals are also present in the lower mass cousins, BH-XRBs e.g. \cite{kara19a} 
This provides further evidence that these features are associated with inner disc-reflection, rather than more distant absorption processes, as suggested by e.g. \cite{milleretal10a}.

\subsection{Quasi-periodic oscillations}
One of the most important results in the study of stellar mass BHs was the discovery of high-frequency quasi-periodic oscillations (HFQPOs) in Galactic BH X-ray binaries, with $\Mbh \sim 5 - 10$\,\Msun. 
The observed Fourier frequencies are typically $\nu_{\rm qpo} \gsim 100-450$\,Hz and are accompanied by a harmonic component in a 3:2 or 2:1 ratio (see e.g. \cite{RemillardMcClintock06,bellonistella14,ingrammotta19}).  They are observed when BH-XRBs are in \emph{Very-high} or \emph{Intermediate} states, characterised by very-high accretion rates.  HFQPOs are the fastest coherent variability signatures observed from BH-XRBs, with frequencies at the \emph{Keplerian} frequency of the innermost stable circular orbit (ISCO).  HFQPOs therefore carry information on the strongly curved spacetime close to the BH, providing constraints on the two fundamental properties of BHs; mass and spin, as well as being important probes of the accretion process.

\begin{figure}[t]
\centering
\includegraphics[width=.6\textwidth,angle=90]{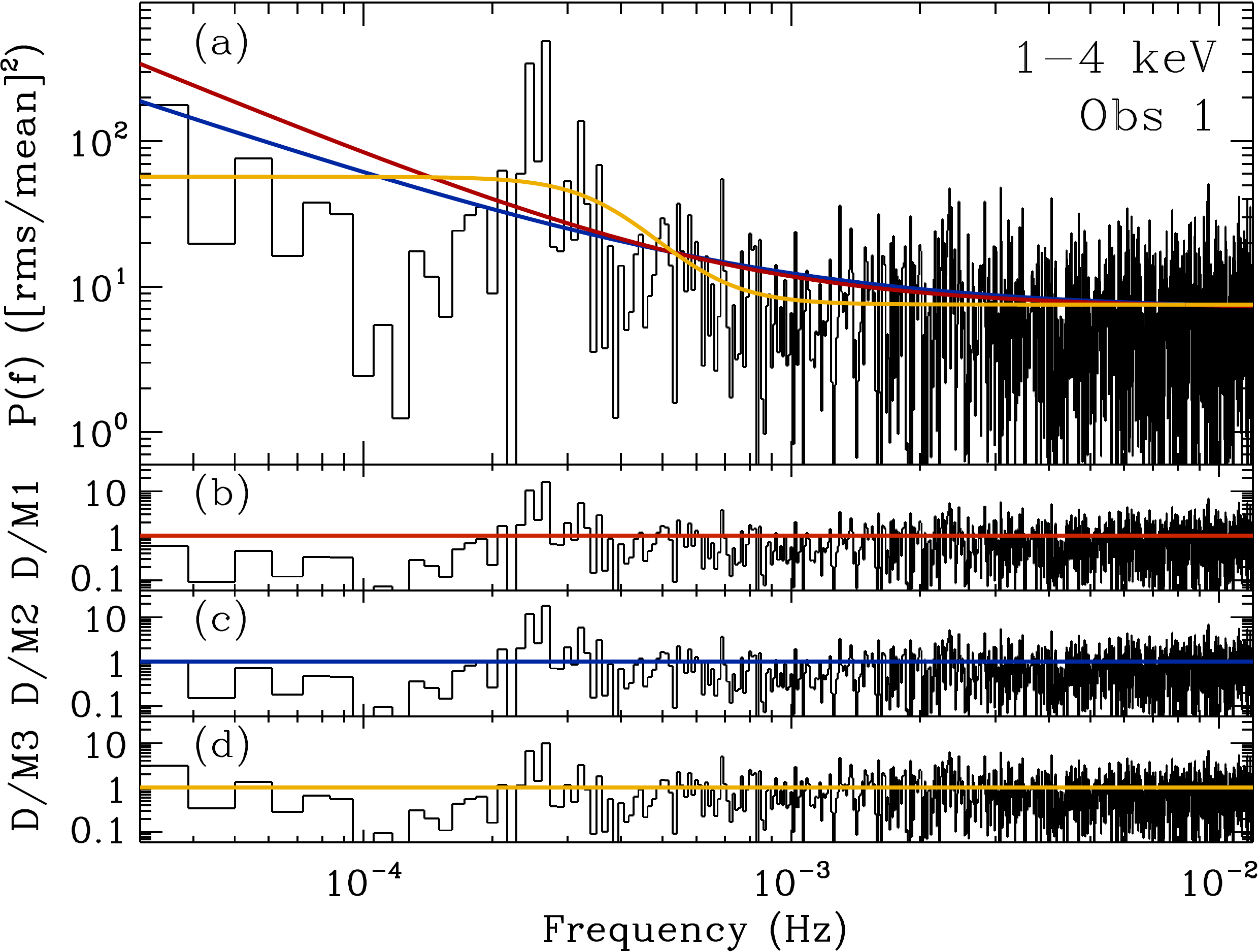}
\caption{\small \rej\ was the first active galaxy to display a significant quasi-periodic oscillation (QPO) in the X-ray PSD \cite{gierlinski08}. The top panel shows the coherent peak in variability power at $2 \times 10^{-4}$\,Hz ($\sim 1$\,hr).  The significance of peaks in power are sensitive to the choice in underlying continuum model. Our lack of knowledge about the true PSD continuum means multiple model shapes should be tested.  The lower panels show the ratio of the data to model.  Figure taken from \cite{alston14a,alston14b}.
}
\label{fig:qpo}
\end{figure}

If general relativity (GR) dominates the accretion process, a scale-invariance implies that HFQPOs should also be present in AGN, with the frequencies scaled as $1/\Mbh$. With $\Mbh \gsim 10^{6-8} \Msun$, we expect $\nu_{\rm QPO} \gsim 1 \times 10^{-4} - 5 \times 10^{-3}$\,Hz (i.e. timescales of $\gsim 200 - 10000$\,s), well within the range of current instruments, such as \xmmn.  Despite having lower count rates than XRBs, the mass ratio means that AGN have higher counts per characteristic timescale, allowing us to study the HFQPO mechanism on a period-by-period basis, with a full suite of both Fourier- and time-domain analysis methods. This is something which can not be done in XRBs with current instruments.

QPOs are notoriously difficult to detect in AGN due to the confusing presence of an underlying broadband (stochastic) noise process.  Early claims of QPO detections were later disfavoured, such as due to confusion with super-orbital period of an off-nuclear ultra-luminous X-ray source.  Other claims were disfavoured for incorrectly modelling the broadband noise continuum when performing significance tests.

A $\sim 1$\,hr periodicity in the Seyfert 1 galaxy \rej\ was the first robust detection of a QPO in an AGN \cite{gierlinski08}.  More recently, \cite{alston14b} showed that the QPO is still present at the same temporal frequency in 5 \xmmn\ observations, and the feature was re-detected in \cite{jin21a}.  Features have been reported in several sources, however \rej\ is the only confirmed QPO in an AGN (e.g. \cite{alston15}).  Despite this importance, the exact mechanism responsible for producing HFQPOs remains unclear, with varying scenarios for the origin of the oscillation presented in the literature (e.g. \cite{wagoner99,stella99,AbramowiczKluzniak01,RezzollaETAL03}).

In the optical band, ground-based monitoring of AGN over the last decade has lead to several claims of periodic light curves (e.g. \cite{graham15a,graham15b,charisi16}).  However, given the low number of period cycles in these data, it is difficult to distinguish these from the typical red noise variability behaviour of AGN (see \cite{vaughan16} for a discussion on these results).

\subsection{Quasi-periodic eruptions} 
\begin{figure}[t!]
\centering
\includegraphics[width=.96\textwidth,angle=0]{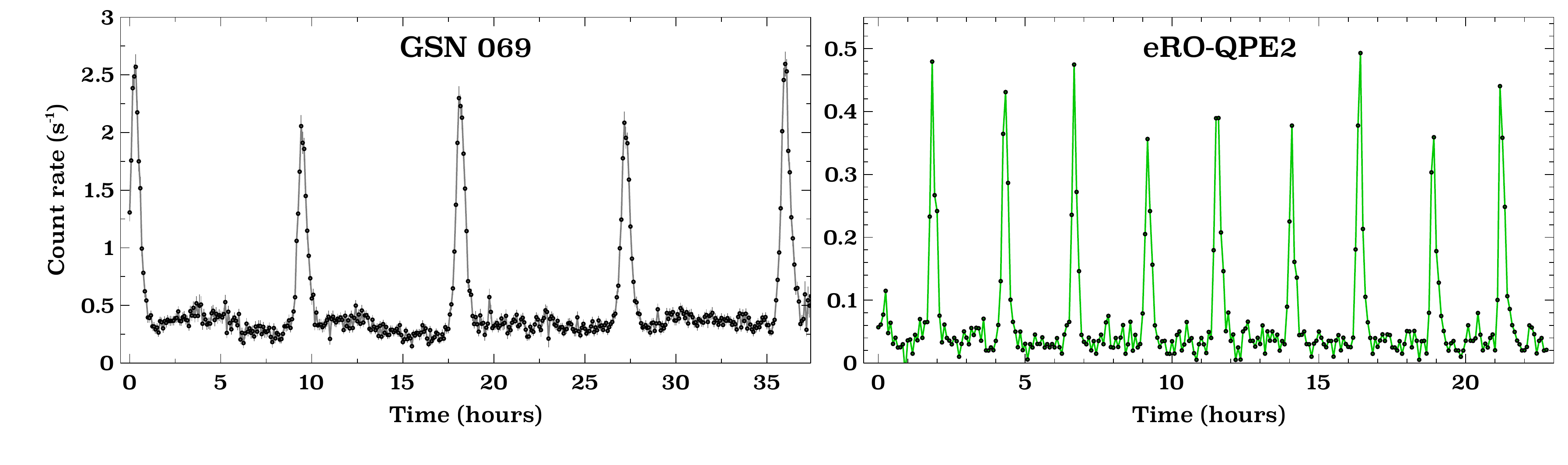}
\caption{\small Two examples of X-ray QPEs observed so far. X-ray light curves from the {\it XMM-Newton}  EPIC-pn instrument extracted in the $0.2-2$ keV band are displayed binned to 300 s. Strong, sharp, and regular X-ray flares recurring every few hours are present in both sources. \textbf{Left panel:} X-ray QPEs observed in GSN 069 in January 2019 (adapted from \cite{2019Natur.573..381M}). \textbf{Right panel:} X-ray QPEs observed in eRO-QPE2 in August 2020 (adapted from \cite{2021Natur.592..704A}). 
}
\label{fig:QPEs}
\end{figure}

X-ray quasi-periodic eruptions (QPEs) are a new cosmic phenomenon  associated to non-stationary mass accretion onto SMBHs. 
Discovered at the end of 2018 in the galaxy GSN 069 \cite{2019Natur.573..381M}, X-ray QPEs are high-amplitude flares of emission of $L_{0.3-2\,keV}\sim 10^{42-43}$ erg s$^{-1}$ over a much stable (quiescent, or plateau) flux level; they last hours or fractions of hours, and recur on time scales of hours. 
The QPE pattern of variability is completely different from the stochastic variability usually observed in AGN (Figure \ref{fig:QPEs}). 
As of October 2021, X-ray QPEs have been observed multiple times in four galaxies: GSN 069 \cite{2019Natur.573..381M}, RX J1301.9+2747 \cite{2020A&A...636L...2G},  eRO-QPE1 and eRO-QPE2 \cite{2021Natur.592..704A}. These last two sources have been discovered during the first 1.5 years of the X-ray all-sky survey by the satellite eROSITA, and more QPEs are expected to be discovered during the completion of the eight-years-long survey \cite{2021Natur.592..704A}.
One and a half QPE-like flares had been also identified in archival {\it XMM-Newton}  data of the star-forming/AGN galaxy J0249, which however by August 2021 had no QPE-like activity anymore \cite{2021ApJ...921L..40C}.
All the QPE-hosting galaxies show signs of optical nuclear activity in excess of starlight \cite{2022A&A...659L...2W}.\\



Spectroscopically, QPEs correspond to oscillations between a colder quiescent  phase with a thermal-like spectrum of $kT\sim 50-70$ eV and a warmer eruptive phase with $kT\sim 100-150$ eV. 
No correlated variability has been detected in the optical or UV bands during X-ray QPEs, making them phenomena restricted to the soft X-ray band and therefore related to the inner accretion flow around the SMBH, being it a disc or another geometry. 
The QPE amplitude is energy-dependent and peaks in the $0.6-1.0$ keV band. Most of the X-ray emission of QPEs  fades away at $E > 1$ keV, and 
the typical hard X-ray power-law emission associated with the X-ray corona is absent or negligible in the QPE-hosting AGN\cite{2019Natur.573..381M,2020A&A...636L...2G,2021Natur.592..704A}.
The QPE-hosting galaxies have low stellar mass and low SMBH mass, of the order of $10^{5-6}\,M_{\odot}$ \cite{2022A&A...659L...2W}. Their nuclear activity seems to be strongly connected to tidal disruption events, thus making QPEs a probe of mass accretion in young SMBH-galaxy systems in real time.\\

The physical interpretation of QPEs is still open, and multiple scenarios, all estimated to take place in the inner regions around the SMBH (from a few $100\,R_g$ down to a few $R_g$), have been proposed, including instabilities of the inner accretion flow (see e.g., \cite{2020A&A...641A.167S,2021ApJ...909...82R}), gravitational lensing of SMBH binaries (e.g., \cite{2021MNRAS.503.1703I}), and orbital scenarios where one or more compact objects of stellar-mass size are orbiting one SMBH (e.g., \cite{2020MNRAS.493L.120K,2021ApJ...917...43S,2021ApJ...921L..32X,2022ApJ...926..101M,2021arXiv210903471Z}).

The last scenario corresponds to extreme mass ratio inspirals (EMRIs), where the evolution of the system is accompanied by strong emission of gravitational waves. This would make QPEs a precursor of the signal in GW detected in the next decade by space-based GW observatories such as Tianqin and LISA (see e.g.  \cite{2021arXiv210903471Z}).
Importantly, in the EMRI scenario, the orbits of the stellar-mass compact object(s) and the trajectories of the X-ray photons emitted during QPEs are strongly influenced by the metric of the spacetime close to the SMBH, which depends on the SMBH spin. It follows that, by accurately measure the time of arrival of X-ray QPEs, the spin of the SMBH could be constrained, in a manner completely independent of spectral modelling (e.g., \cite{2021ApJ...921L..32X}).


\subsection{Optical/UV variability}
\label{ouvvar}


Optical variability in quasars has been known for several decades (e.g. 3C 273; \cite{Press1978}).  Many Seyfert galaxies show both UV and optical variability, with an rms amplitude of 1-10\%.  On faster timescales, a weaker variability amplitude than that of the X-rays is observed (see bottom panel of Fig.~\ref{fig:psd}).  However, the origin of the optical and UV variability is still unclear: is it caused by the inward propagation of fluctuations in the disc? via reprocessing emission from some other band? Or both?\\

The location of the optical/UV emitting region depends on the details of the accretion flow, and these in turn depend on the BH mass and accretion rate, but is typically $\sim 10-1000~\Rg$.  The causal connections between these processes can be investigated by studying the time variations in the luminosity across different wavebands (see also Sect. \ref{correlradioUVX} for the observed UV-X-ray correlation in sample of AGN).  If the emission mechanisms are coupled, the optical/UV and X-ray variations should be correlated, in which case the direction and magnitude of time delays should reveal the causal relationship.  Two favoured coupling mechanisms are:

\begin{enumerate} \itemsep1pt \parskip0pt
  \item \emph{Compton up-scattering} of optical/UV photons --- produced in the disc --- to X-ray energies in the corona \cite{HaardtMaraschi91}.  Here, any variations in the seed photons being upscattered will directly translate into variations in the X-ray emission.
  \item \emph{Thermal reprocessing} in the disc of X-ray photons produced in the corona \cite{guilbertrees88}. Reprocessing in a cold disc is observed in X-ray reflection features of AGN spectra (e.g. \cite{nan94,Fabian2002kdblur}). The wavelength the reprocessed photons emerge at and their contribution to the energy radiated by the disc can be tested in multi-wavelength variability studies.  Of course, X-rays will irradiate a significant portion of the disc, but we are only interested in the region --- or annulus --- of the disc that reprocesses the X-rays photons to UV and optical energies.
  A standard  'Shakura \& Sunyaev' accretion disc \cite{shaksuny73} with temperature profile $T \propto R^{-3/4}$ will give rise to wavelength-dependent lags, $\tau$, following $\tau \propto \lambda^{4/3}$, e.g. \cite{cackett07}.  Optical/UV continuum light curves have long been known to be well-correlated, and to have short interband lags of a few days, at most. 
\end{enumerate}

\noindent The interaction timescale for these two processes is approximately the light crossing time between the two emitting regions, and will be in the region of minutes to days for BH masses $\Mbh \sim10^{6}-10^{8} \Msun$.\\

 
Studies of correlations between variations in different wavebands are a potentially powerful tool for investigating the connections between different emission mechanisms (see \cite{uttley06} for a short review).  \cite{done90} performed the first simultaneous optical/X-ray study, searching for correlated emission in the NLS1 galaxy NGC 4051 over two days.  The X-ray flux varied by a factor of two over this period, whilst the optical variations were $\lsim 1$\%.  This demonstrated that optical and X-ray emission was produced by separate processes.  X-ray/optical correlations on long timescales have been seen in radio-quiet AGN (e.g. \cite{uttley03,arevalo08,arevalo09,breedt09}).  Together with the optical-optical lags (e.g. \cite{cackett07}), they imply that a combination of accretion fluctuations and reprocessing produces much of the optical variability.  


More recently, high-cadence observations with multiple snapshots per day, combining {\it Swift}  and ground-based observations, have made significant improvements to the continuum lag measurements \cite{edelson15,edelson19,mchardy18,cackett18}, see \cite{cacket21rev} for a review.  These deep multi-instrument campaigns have found results common across several AGN: i) that the lags approximately follow the $\tau \propto \lambda^{4/3}$ relationship; ii) the normalization of this relation is larger than expected from standard accretion disc theory; iii) the lags in the U band are systematically longer, of order a factor of 2 than expected based on an extrapolation of the above relation fit to other optical and UV bands (e.g. \cite{edelson19} but see \cite{kammoun21}). 

Using HST spectroscopic monitoring of NGC 4593, \cite{cackett18} resolved the continuum lags, showing a clear discontinuity at the Balmer jump.  These results can be explained if there is an additional continuum emission component to the lags which do not come from the disc, but from the broad line region (BLR; e.g. \cite{koristagoad2001,lawther18}). This diffuse continuum from the BLR emits across the full UV/optical bands, and can affect the lags in all wavelengths. The time lags from the BLR components also increases with increasing wavelength, except for the discontinuities at the Balmer and Paschen jumps.  This makes it difficult to cleanly separate the BLR lags from the disc lags \cite{koristagoad2001}; however, the different size scales of the two components means these two lagging components should in principle be separable by studying the signals on different timescales: with disc reverberation taking place on timescales of a few days and BLR continuum reverberation taking place on timescales of weeks or longer. This is currently an active area of research, which will provide insight into the inner accretion flow (e.g. \cite{kammoun21,cackett21}).

%

\section{Accretion properties in AGN populations: the disc-corona coupling}
\label{correlradioUVX}
As explained in Sect. \ref{corona}, the X-ray emission is expected to come from a small region close to the BH, the so-called hot corona, while the UV emission is produced by the inner part of the accretion disc, also in the vicinity of the BH (see Fig. \ref{fig:AGNsketch}). It is generally believed that the X-rays are produced by Comptonisation of the UV photons by the hot electrons of the corona. 
Both emission, in UV and X-rays, are then expected to be linked one with each other and, indeed, correlation on timescale ranging from minutes to days, depending on the BH mass, are observed (see Sect. \ref{ouvvar}). The understanding of the mutual behaviour between X-rays and UV should give then important information on the corona-disc structure. \\
\begin{figure*}[t]
\centering
\includegraphics[width=\textwidth,angle=0]{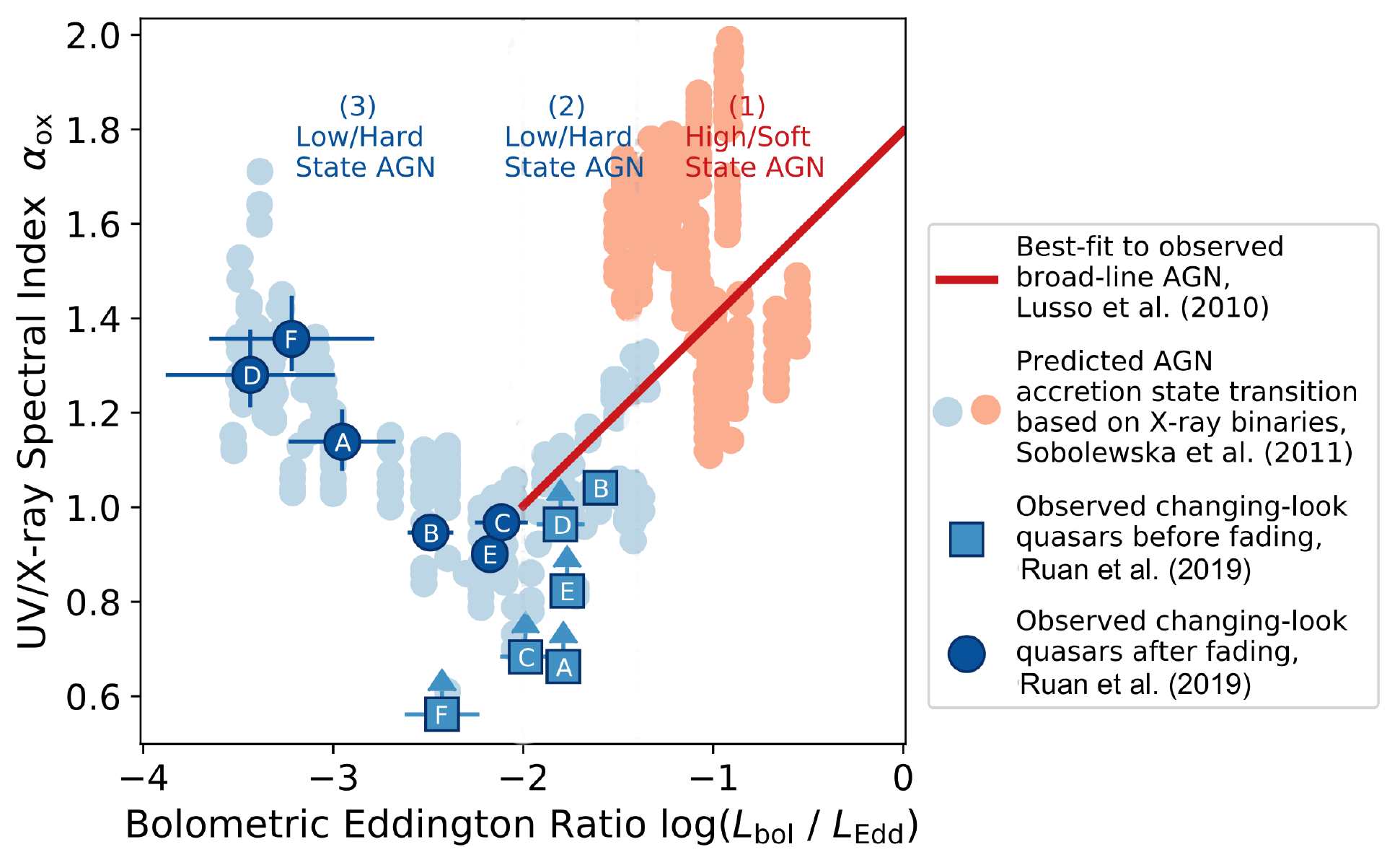}
\caption{
Evolution of the $\alpha_{ox}$ with luminosity: at low luminosity, the so-called changing-look quasars before (blue squares) and after (dark blue circles) fading; at high luminosity, the correlation observed in the sample of bright quasars used in the top plot is reported as a red line. These observations are compared with observations of X-ray binary transitions from a high/soft state (light red circles) to a low/hard state (light blue circles). The similarity in the spectral behavior of AGN and X-ray binaries suggests that the geometries of BH accretion flows behave in a similar way with the bolometric luminosity. See \cite{rua19} for more details. \label{alphaoxlbol}}
\end{figure*}

While UV vs X-ray correlations are observed from object to object and inform us about the intrinsic behavior of each AGN, interesting correlations are also present when looking to samples of AGN. Both types of correlations are complementary, since the latter put constraints on the global behavior of the accretion process in AGN as a group rather than separated individuals. \\

Due to the absorption from our own Galaxy, mainly due to the neutral hydrogen, we have no access to extragalactic photons with energy between $\sim$10 eV and a few hundreds of eV. The continuum between the Optical/UV and X-rays is then often parameterized by a simple power-law whose spectral index 
$\alpha_{ox}=-\log (L_{2 keV}/L_{2500 \AA})/\log (\nu_{2 keV}/\nu_{2500 \AA})$\footnote{We follow the definition by \cite{tan79}. With this definition the observed $\alpha_{ox}$ is generally positive.}, also called the optical-to-X-ray spectral index, is computed using the rest-frame luminosities $L_{2 keV}$ and $L_{2500 \AA}$ at 2 keV and 2500 \AA\ respectively. The value of $\alpha_{ox}$ gives then an idea of the importance of the UV emission with respect to the X-ray emission.

A non-linear correlation is known since decades, when looking to sample of AGN, between $\alpha_{ox}$
and the logarithm of $L_{2500\AA}$, i.e. $\alpha_{ox}=A\log L_{2500\AA} + B$ with A=$0.1-0.2$. A similar correlation is also observed with the logarithm of the bolometric luminosity in Eddington units, the so called Eddington ratio $\lambda_{Edd}=L_{bol}/L_{Edd}$ (see Fig. \ref{alphaoxlbol}), e.g. \cite{avn82,wil94,str05}).
Another non-linear correlation is also observed between the X-ray and UV luminosities, $L_{2 keV}\propto L_{2500 \AA}^p$ with $p\sim$0.6-0.8 (e.g. \cite{stef06,just07,lus20}). 
\begin{figure*}[t]
\includegraphics[width=\textwidth,angle=0]{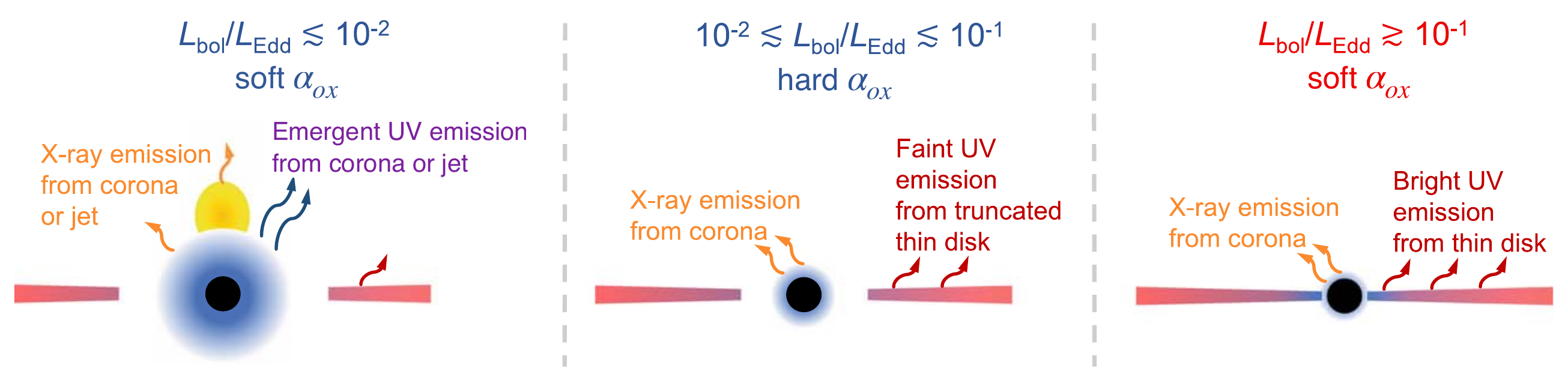}
\caption{Possible evolution of the geometry of AGN accretion flows to explain the change of the $\alpha_{ox}-L_{UV}$ correlation at high and low Eddington ratio (see Fig. \ref{alphaoxlbol}). At high Eddington ratios (right panel; $L_{bol}/L_{Edd}\geq 0.1$), the high values of $\alpha_{ox}$ are due to the bright UV emission from a thin accretion disc present down to the ISCO. As the Eddington ratio (middle panel) drops ($0.01\leq L_{bol}/L_{Edd} \leq 0.1$), the inner regions of the thin disc become progressively truncated. The truncation of the inner disc causes the UV luminosity to fade, and thus $\alpha_{ox}$ hardens (i.e. decreases). At low Eddington ratios ($L_{bol}/L_{Edd}\leq 0.01$, left panel), the $\alpha_{ox}$ softens/increases again due to the emergence of UV emission from the hot inner accretion flow or the jet. From \cite{rua19}. \label{AGNscheme}}
\end{figure*}
These different non-linear correlations becomes even tighter when effects like intrinsic UV/X-ray variability is taken into account (e.g. \cite{Vagnetti13,Chiaraluce18}). They all bring the same information and tell us the way the UV-X-ray SED of AGN evolves with the disc luminosity and the global luminosity of the system. They suggest that the X-ray corona is relatively less luminous for high-luminosity accretion disc compared to low-luminosity cases (see, e.g., \cite{kub18,arc19} for different possible scenarii). In other words, the more luminous the AGN, the more dominant the emission of the accretion disc with respect to the hot corona. These non-linear correlations are observed up to high redshift (e.g. \cite{ris19,velt20,zhen20} and see Chap. 4, this Volume).\\

The slopes of these correlations may, however, depends on the luminosity of the sources. For instance, for bright sources (i.e. $\lambda_{Edd}>1\%$), a positive correlation between $\alpha_{ox}$ and $L_{2500\AA}$ or $\lambda_{Edd}$ is generally observed (e.g. \cite{stef06,gree07,lus10}) 
up to high redshifts (e.g. \cite{nan17,salv19}). 
At lower luminosity ($\lambda_{Edd}<1\%$), however, there are hints of an anti-correlation (see bottom of Fig. \ref{alphaoxlbol}, \cite{rua19}). 
This could indicate a change of the radiative properties of the emitting source and/or the geometry of the disc-corona system between high and low luminosity states (see e.g. \cite{sob11}). For instance, for bright AGN, the accretion disc may extend close to the BH, letting very few power to the X-ray corona. On the contrary,  low-luminosity AGN may have a truncated disc which radiates less in UV, while the bright X-ray corona occupied a large part of the inner regions close to the BH (see e.g. Fig. \ref{AGNscheme}).\\

\begin{figure*}
\begin{center}
\includegraphics[width=0.8\textwidth,angle=0]{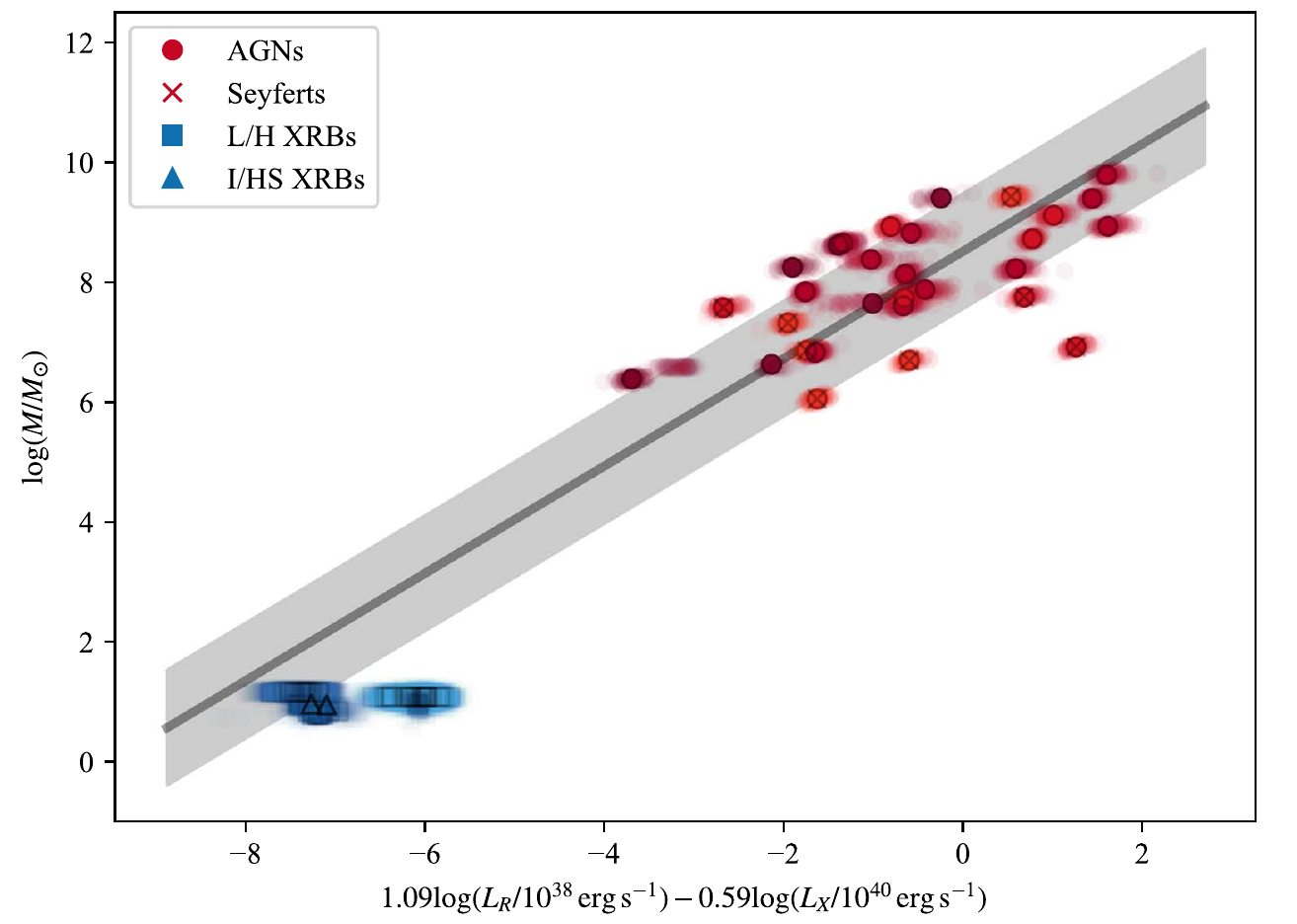}
\end{center}
\caption{The fundamental plane of BH accretion, an empirical correlation of the mass of a BH (M), its radio continuum luminosity ($L_R$), and its X-ray luminosity ($L_X$), from \cite{gul19}. Red points are for AGN and blue for XRBs, the shade is given by the value of $\log(L_X/L_{Edd})$. The best-fit relation is plotted with a dark gray line with a light gray shaded region to indicate the 1$\sigma$ region of the Gaussian intrinsic scatter.\label{fondplane}}
\end{figure*}

While it was believed that this conclusion applied only for radio-quiet (RQ) AGN, whose disc-corona emission is not ``polluted" by the emission from the jet, a similar non-linear correlation, with a similar slope, has been observed also in radio-loud (RL) AGN (e.g. \cite{zhu20}). 
This results suggests a common coronal origin for the X-ray emission of both types of objects (only a small proportion of radio-loud AGN, the so-called flat-spectrum radio quasars, could be dominated by a distinct jet component in the X-rays, see e.g. \cite{zhu20}. See also Chap. 1 this Volume).\\ 

It is true, however, that 
the X-ray luminosities of RL AGN are systematically larger than RQ AGN, and the difference increases with the radio loudness $R= L_{5 GHz} /L_{4400 \AA}$ (where $L_{5GHz}$ and $L_{4400 \AA}$ are the monochromatic luminosities at rest-frame 5 GHz and 4400 \AA, respectively \cite{kel89}). 
Since the radio emission is generally interpreted as a jet signature, these results suggest also a link between the jet and the corona emission processes. This is also supported by the non-linear correlation observed between the radio luminosity $L_{R}$ and the X-ray luminosity $L_X$ in compact objects, either in X-ray binaries (e.g. \cite{cor00,coriat11}) 
or in AGN (the so-called fundamental plane of BH activity, see Fig. \ref{fondplane} e.g. \cite{mer03}).  
Overall, the observed relations between $L_{UV}$, $L_{X}$ and $L_{R}$, while not fully understood, are clear evidence for a connection between the jets and the disc/corona structures.







%

\section{Future prospects}
\label{futur}
The observational context in a near and far future in X-ray astrophysics is quite promising. The first X-ray polarisation mission ({\it IXPE}), since the first X-ray polarisation detection of the Crab nebula in the 70's (e.g. \cite{wei76}), was launched in Dec. 2021 \cite{ixpe} and should provide very important constraints on the close environment of compact objects (see Sect. \ref{xraypolar}). Other X-ray polarimetry missions, e.g. {\it eXTP} (enhanced X-ray Timing and Polarimetry \cite{eXTP}), planned to be launched in 2027, or {\it XL-Calibur}, a balloon experiment planned to be launched in 2023 \cite{xlcalibur}, should be available in the next years. 

The X-ray spectroscopy domain will be soon revolutionized by the availability of X-ray microcalorimeter observations on both the {\it XRISM} and {\it Athena} missions (see Sect. \ref{microcal}). These observations will provide a uniform spectral resolution of only a few eV over a broad energy range (typically $0.2-12$ keV), and will represent a major step in our understanding of the matter properties around compact objects, as the very few observations of {\it Hitomi} were able to demonstrate during the short life of the mission \cite{hit18,2016Natur.535..117H}.
In 2023, {\it XRISM} (the X-ray Imaging and Spectroscopy Mission) will be launched, and its soft X-ray spectrometer with microcalorimeter technology will improve the spectral resolution by more than a factor of 10 with respect to the present best X-ray spectrometers.
Last but not least, {\it Athena} (the Advanced Telescope for High ENergy Astrophysics), selected by ESA within the Cosmic Vision Program, will be the next cornerstone mission of the X-ray astronomy domain for the following decade \cite{nan13} and should be launched in the 30's, providing a further substantial improvement over the {\it XRISM} performances. 

\subsection{X-ray polarimetry}
\label{xraypolar}
Close to the BH, which are the regions of interest for this chapter, the disc-corona structure is the major source of the polarisation signal. As said previously, the hot coronal X-ray emission  (which is the dominant one in the 2-10 keV energy range) is likely due to Comptonisation of the disc photons in the hot corona, and it is therefore expected to be polarised (e.g. \cite{haardtMatt93,pou96}). The polarisation degree and angle actually depend  on several physical parameters of the disc-corona system, like the temperature and optical depth of the hot corona, the spin of the BH, the scale height of the corona, these effects being strongly energy dependent (e.g. \cite{sch10}).

Other sources of polarisation, in addition to Comptonisation, are also expected, especially from reflection of the X-ray emission on the matter surrounding the SMBH. For instance, part of the coronal X-ray emission is expected to be intercepted and reflected by the accretion disc itself, giving rise to the Compton Reflection component discussed in Sect. \ref{reprocessSect}. This component, which becomes relevant above 5 keV, is also expected to be polarised, the polarisation degree depending mainly on the inclination angle of the disc (e.g. \cite{mat93}). Moreover, due to the curved space-time around BHs, the polarisation direction of the X-rays emitted close to the event horizon rotates in a way that depends on the location of the emitting point with respect to the BH and the spin of the BH (e.g. \cite{dov11}). The X-ray polarisation observations to come could then help to put constraints also on the BH spin. \\

Compared to galactic BH systems, however, polarisation measurement are expected to be quite challenging in AGN mainly due to their lower X-ray flux which is, for nearby Seyfert galaxies and quasars, at least an order of magnitude smaller than the brightest galactic BHs. This weakness can be partly compensated by the lower temperature of the AGN accretion disc. Indeed, more scattering events are then required to inverse Compton scatter these seed photons to X-ray energies $>$2 keV, naturally increasing the degree of polarisation of the X-ray signal. Moreover, a cooler disc also means larger absorption opacities. The range of available scattering angles for X-ray photons to escape the disc-corona systems is then more restricted. Together, these features, which are particular to SMBH environment of AGN, should lead to a stronger ability to discriminate between different disc-corona geometries and properties (e.g. \cite{sch10,kra12,beh17b,tam18}). This is crucial since, as explained in Sect. \ref{corona}, constraints on the geometry are not easily obtained from spectral or timing results given the degeneracy of the models generally used.\\ 

\begin{figure}
\begin{tabular}{c}
\includegraphics[width=\textwidth]{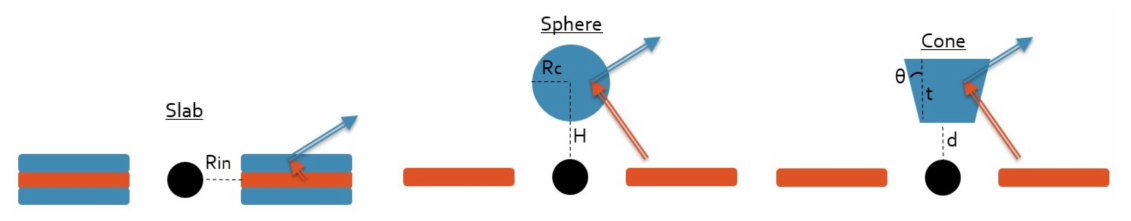}
\vspace*{1cm}\\
\includegraphics[width=0.8\textwidth]{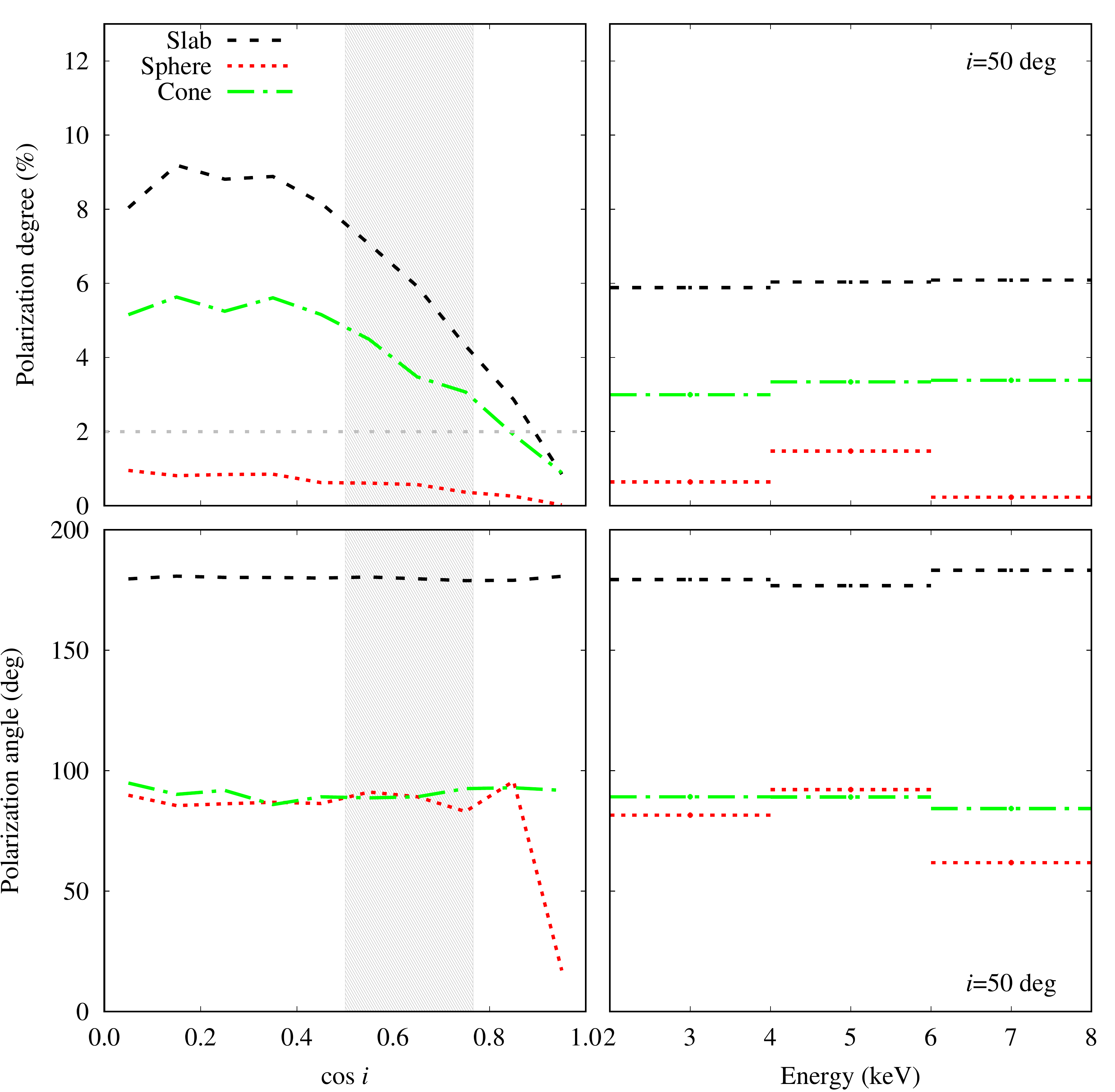}
\end{tabular}
\caption{{\bf Top:} Different disc-corona geometries. The X-ray emission from the corona (in light blue) is produced by inverse Compton scattering of the optical-UV radiation from the disc (in orange). {\bf Bottom:}  Comparison between the (2-8 keV) polarisation degrees (upper plots)  and the polarisation angle (lower plots) for the slab, spherical, and conical geometry, versus the cosine of the inclination angle of the observer (left) or the energy (right). In these simulations the X-ray emission is characterized by a photon index of 1.8 and a coronal temperatures of 50 keV. The shaded area corresponds to the inclination interval 40-60 deg, while the gray dotted line shows the minimum polarisation detectable by {\it IXPE}  in a 500 ks exposure of the AGN MCG-5-23-16 \cite{Balokovic15,zoghbi17}. The polarisation degree is always larger for the slab. An MDP of $\sim$2\% should be sufficient to discriminate the different geometries from each other.  From \cite{urs22}. }
\label{fig:polar}
\end{figure}

This is exemplified in Fig. \ref{fig:polar} which compares the expected X-ray polarisation degree and angle for different disc-corona geometries. Although the different geometries can produce similar X-ray spectral shape (but with different coronal temperature  and/or optical depth, see Fig. \ref{fig:degeneracy}), the stronger their asymmetry the larger the X-ray polarisation signal. A slab-like geometry will then have a stronger polarisation signal compared to a sphere-like one, as shown in the right plots of Fig. \ref{fig:polar}. Consequently, by measuring the polarisation degree in a sample of AGN with different inclination angles, we should be able to constrain the most favorable geometry and discriminate between different scenarios. Reversely, for objects for which we can have  independent estimates of the disc inclination (e.g., Fe line fitting), the amplitude of polarisation in the 1-10 keV band can be used to determine the disc-corona structure. A few bright AGN should be observed during the {\it IXPE}  mission and a Minimum Detectable Polarisation (MDP\footnote{The MDP is the degree of polarisation detected at the 99\% confidence level independent of the position angle.}) of 2\% should be sufficient to discriminate between these geometries. This would require, however, exposures of several hundreds of thousands of seconds, depending on the AGN luminosities.  In a more distant future, {\it eXTP} should go a step further allowing to measure polarimetry signal in the weakest AGN, thanks to the larger effective area, and by enabling simultaneous
spectral-timing  studies of cosmic sources in the energy range
from 0.5-30 keV.\\

Other reflecting regions at larger scales (i.e. $> 1000 R_g$) are present in the environment of radio-quiet AGN and are expected to produce a polarimetric signal, like the so-called dusty ``torus" envisaged in Unification models (e.g. \cite{antonucci93}) or the polar outflowing winds and warm absorber region (e.g. \cite{Halpern84}). 
All in all, polarimetric observations are thus expected to be of precious help to constrain the global geometry of these different components as well as their basic properties (column density, inclination, etc...) among different AGN types (e.g. \cite{goo11b,marin18a}). \\

\subsection{X-ray microcalorimeters: XRISM, Athena}
\label{microcal}


Our view of the inner regions of AGN will be sharpened by the advent of X-ray spectrometers with microcalorimeter technology, such as {\it Resolve} onboard {\it XRISM} \cite{2020arXiv200304962X} and {\it X-IFU} onboard {\it Athena} \cite{2018SPIE10699E..1GB}. These spectrometers will provide non-dispersive, high-resolution X-ray spectra with a uniform resolution over a wide energy band. 
Importantly, they will open a window on the Fe K band of AGN observed at high resolution with high photon counting statistic. This will allow to break the degeneracy between reflection-dominated and absorption-dominated models for the inner regions of AGN (Sect. \ref{sec:reprocessing}).

\begin{figure}
\begin{center}
\includegraphics[width=1.\textwidth,angle=0]{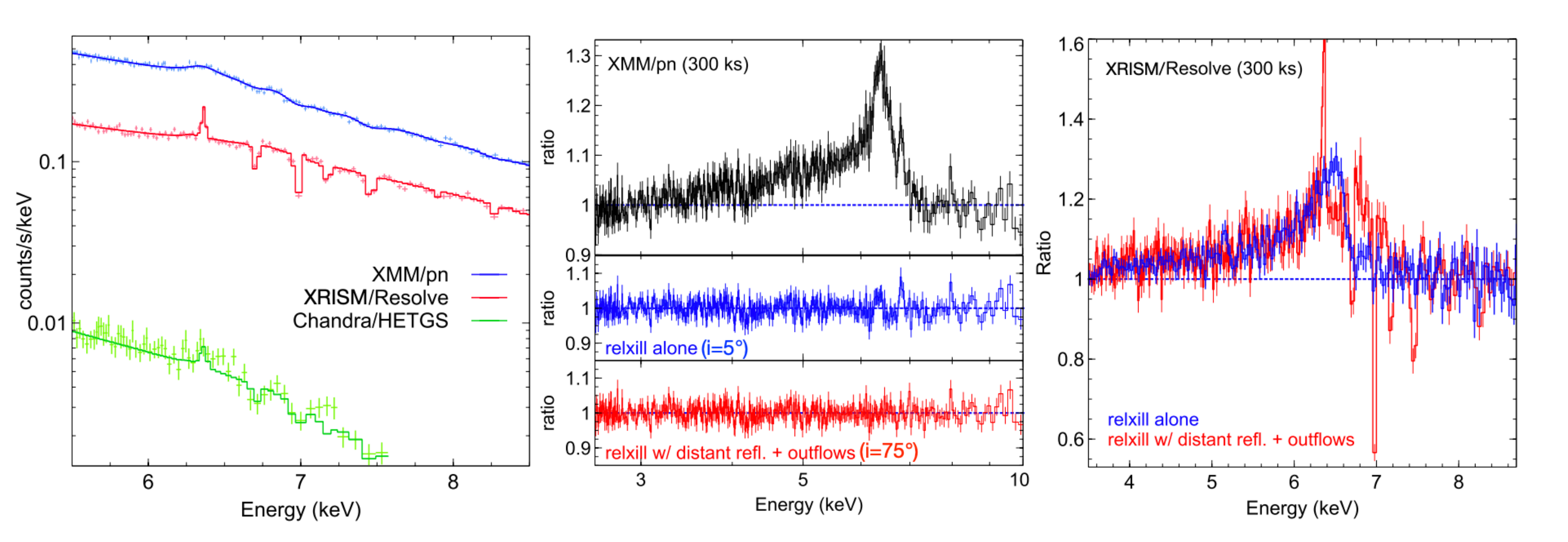}
\end{center}
\caption{{\bf Left panel:} comparison of the spectral view of the Fe K band of the Seyfert galaxy MGC-6-30-15 observed for 300 ks by {\it XMM-Newton}, {\it Chandra}, and {\it XRISM}. {\bf Central panel}: the Fe K line profile observed by {\it XMM-Newton} (black) can be equivalently well fitted with a relativistic reflection model (blue) and with distant reflection plus outflows (red). {\bf Right panel:} spectral simulation with {\it XRISM/Resolve} with marked differences between the reflection-dominated scenario (in blue) and the wind-dominated scenario (in red).  
From \cite{2020arXiv200304962X}.\label{xrism}}
\end{figure}

A demonstration of the expected breakthroughs provided by microcalorimeter observations is shown in Fig.~\ref{xrism} for the case study of the bright Seyfert galaxy MCG-6-30-15. In the left panel, the Fe K spectral region of MCG-6-30-15 observed by the CCD-resolution EPIC-pn instrument onboard {\it XMM-Newton} is shown in black, the same region observed by the high-resolution {\it HETG} onboard {\it Chandra} is shown in green, and the expected {\it Resolve} observation is simulated in red. In the central panel, the Fe K spectral region of MCG-6-3-15 observed by {\it XMM-Newton} is shown in black as ratio between the data and a simple power-law model; the two smaller lower panels show spectral residuals in the case of a fit to a relativistic reflection model (in blue) and to a distant reflection with outflows model (in red): the two models are statistically indistinguishable with the currently available data. The right panel of Fig.~\ref{xrism} shows two 300 ks-long {\it XRISM} spectral simulations of MCG-6-30-15 for the two physical scenarios above: the high-resolution of the {\it Resolve} microcalorimeter will definitely permit to disentangle the contribution from distant reflection from the relativistic one, and to clearly detect the signatures of outflowing gas, if present, as a series of blueshifted absorption lines.

While the moderate effective area of {\it XRISM} will allow a detailed study of the Fe K region of only the brightest AGN, many more sources will be observable at high-resolution with the {\it X-IFU} microcalorimeter onboard {\it Athena}. This will have comparable or better spectral resolution than {\it Resolve} and a much larger effective area, thus bringing the detailed knowledge of the reprocessing features around SMBHs in both weak and distant AGN. 

The large effective area of the instruments onboard {\it Athena} will be crucial also for timing studies of AGN.
The power of X-ray reverberation measurements is that they are able to map the emitting regions close to compact objects in terms of the absolute physical scale, rather than in terms of $R_g$ achievable with spectral fitting alone. The scales that can be mapped are sub-micro to nano-arcseconds in angular size on the sky - smaller than can ever be accessed with any X-ray imaging technology, for decades into the future. Large detected count rates are key, especially for XRBs where we have seen that signal-to-noise scales linearly with count rate. Thus the most important requirement to access these scales is a large effective area instrument, capable of measuring large count rates, combined with good energy resolution (CCD-quality or better) to probe the variable relativistically-smeared reflection.

Significant breakthroughs which exploit the full potential of X-ray reverberation will require a step up to square metres of collecting area, which will be attained by the end of the 2020s. The {\it Athena} mission \cite{nan13} will increase collecting area compared to {\it XMM-Newton}  by more than an order of magnitude at soft X-rays, and by a factor of 3-4 at Fe K energies, enabling significantly improved X-ray spectral-timing and reverberation measurements for both AGN, as well as XRBs. 

The sensitivity of {\it Athena} to faint sources, especially in the soft band, will allow the reverberation signal to be discovered in many more fainter objects, so reaching to a much wider luminosity range of sources than is accessible today and opening up the study of the innermost regions of a wide variety of AGN classes.



\bibliographystyle{spphys}

\end{document}